\pdfoutput=1
\PassOptionsToPackage{hyphens}{url}
\documentclass[pdflatex,sn-mathphys-num]{sn-jnl}

\usepackage{graphicx}%
\usepackage{multirow}%
\usepackage{amsmath,amssymb,amsfonts}%
\usepackage{amsthm}%
\usepackage{mathrsfs}%
\usepackage[title]{appendix}%
\usepackage{xcolor}%
\usepackage{textcomp}%
\usepackage{manyfoot}%
\usepackage{array}
\usepackage{booktabs}%
\usepackage{algorithm}%
\usepackage{algorithmicx}%
\usepackage{algpseudocode}%
\usepackage{listings}%
\usepackage{tabularx}%
\usepackage{makecell}%
\usepackage[nointegrals]{wasysym}%

\newcommand{\fullsupp}{\CIRCLE}
\newcommand{\partsupp}{\LEFTcircle}
\newcommand{\nosupp}{\Circle}

\definecolor{yamlkey}{HTML}{0B5394}
\definecolor{yamlbool}{HTML}{A626A4}
\definecolor{yamlcmt}{HTML}{7F8C8D}
\definecolor{yamlstr}{HTML}{6A8759}
\lstdefinelanguage{yaml}{
  morekeywords=[1]{network,transit,demand,simulation,horizon,value,unit,%
    defaultSpeedLimitKph,saturationFlowVehPerHourPerLane,%
    defaultDwellTime,iteration,enabled,maxIterations,%
    convergenceWindow,convergenceMeanTolerance,weightUpdateMethod,%
    replanning,routeProbability,modeProbability,departureTimeShiftProbability,scoring,%
    marginalUtilityOfTravelTimePerHour,modeConstants,car,%
    output,directory,formats,source,path,lines,schedules,stops,%
    geoparquet,randomSeed,jamSpacingM,schema_version,name,version,%
    time_unit,models,variables,entities,attributes,blocks,entity,%
    rate,mean,to,delay,min,mode,max},
  morekeywords=[2]{true,false,null,yes,no},
  keywordstyle=[1]\color{yamlkey}\bfseries,
  keywordstyle=[2]\color{yamlbool}\bfseries,
  sensitive=true,
  comment=[l]{\#},
  commentstyle=\color{yamlcmt}\itshape,
  stringstyle=\color{yamlstr},
  morestring=[b]',
  morestring=[b]",
}

\theoremstyle{thmstyleone}%
\theoremstyle{thmstyletwo}%

\theoremstyle{thmstylethree}%

\begin{document}

\title{A Model-Centric DevOps Architecture for DEVS-Based Digital Twin Simulation Services}


\author*[1]{\fnm{Arnis} \sur{Lektauers}}\email{arnis.lektauers@rtu.lv}
\equalcont{These authors contributed equally to this work.}

\author[1]{\fnm{Gusts} \sur{Linkevičs}}\email{gusts.linkevics@rtu.lv}
\equalcont{These authors contributed equally to this work.}

\author[1]{\fnm{Guntis} \sur{Mosāns}}\email{guntis.mosans@rtu.lv}

\author[1]{\fnm{Arina} \sur{Fokina}}\email{arina.fokina\_1@rtu.lv}

\author[2]{\fnm{Rasa} \sur{Gulbe}}\email{rasa.gulbe@datigroup.com}

\affil*[1]{\orgdiv{Institute of Information Technology}, \orgname{Riga Technical University}, \orgaddress{\street{Zunda Embankment 10}, \city{Riga}, \postcode{LV-1048}, \country{Latvia}}}

\affil[2]{\orgname{Dati Group Ltd.}, \orgaddress{\street{Balasta dambis 80A}, \city{Riga}, \postcode{LV-1048}, \country{Latvia}}}

\abstract{Digital twin simulation models evolve and are redeployed like software. Engines based on the Discrete Event System Specification (DEVS) have a sound formal basis but offer little support for versioning, automated validation, or continuous delivery in cloud-native environments. This paper proposes a model-centric DevOps architecture that deploys DEVS-based digital twin simulations as managed services. Models are first-class DevOps artefacts written in a declarative language based on YAML (YAML Ain't Markup Language) with a tuple-level mapping to the multiPDEVS formalism. Repository continuous integration (CI) validates publishing manifests, the engine loader performs syntactic, structural, and static semantic checks before every simulation, and published model artefacts are versioned and immutable, so that an earlier validated version can be selected by pinning its identifier. The platform runs as containerised microservices on Kubernetes with engine adaptations that externalise inputs and results and expose lifecycle control. An initial case study on bus route~22 in Riga exercises the lifecycle on an offline model with roughly 115{,}000 DEVS atomic components and reports single-container throughput and concurrent job execution on one Kubernetes node. Behavioural validation and multi-node evaluation remain future work.}


\maketitle

\section{Introduction}\label{sec:introduction}

Digital twins are now used across engineering domains for continuous monitoring, prediction, and optimisation of physical systems through their virtual counterparts~\cite{Grieves2017,Tao2019}. Most implementations rely on dynamic simulation to let the virtual model reproduce and predict system behaviour under varying conditions. Among formal frameworks, the Discrete Event System Specification (DEVS)~\cite{Zeigler2018} offers a systems-theoretic basis for modelling such systems.

DevOps practices have reshaped software delivery through continuous integration and deployment (CI/CD) pipelines~\cite{Bass2015}, yet these practices remain insufficiently integrated into the lifecycle of DEVS simulation models. Simulation models evolve iteratively like software, but lack the standardised mechanisms for automated validation, versioned deployment, and rollback that DevOps pipelines provide elsewhere.

Many digital twin implementations are still built ad hoc~\cite{Niyonkuru2021}, and most DEVS engines were built for desktop or cluster execution with persistent processes, direct file system access, and manual configuration. These assumptions clash with cloud-native deployment, where containers are ephemeral, scaling is horizontal, and infrastructure is declared as code.

The research problem addressed in this paper is therefore the absence of an engineering lifecycle in which a DEVS model is specified, validated, versioned, deployed, and reverted through the automated mechanisms that software artefacts already have, while keeping the formal semantics that motivate DEVS. Three research questions follow from this problem and structure the paper:
\begin{enumerate}
\item Can DEVS models be specified declaratively in a form that allows well-formedness with respect to the formalism to be checked before a simulation is executed, enabling validation as a continuous integration (CI) stage?
\item What adaptations enable a classical DEVS engine based on persistent processes to execute as an ephemeral, horizontally scaled cloud workload while ensuring that each run is reproducible from versioned artefacts?
\item Does the resulting lifecycle let a digital-twin model of realistic scale be varied through configuration changes alone, without modifying the engine code, while achieving sufficient throughput for iterative use?
\end{enumerate}

This paper presents a model-centric DevOps architecture for deploying DEVS-based digital twin simulations as managed services. The contributions are the following:

\begin{enumerate}
    \item A declarative Simulation Model Definition Language (SMDL) based on YAML (YAML Ain't Markup Language) for DEVS simulation models with a tuple-level mapping to the multiPDEVS formalism~\cite{Foures2018}, and a companion Simulation Scenario Definition Language (SSDL) that binds model versions to scenario parameterisations, enabling version control, automated validation, and CI/CD integration.
    \item A microservices-based cloud-native architecture with engine adaptations that externalise inputs and results and expose lifecycle control, letting a classical DEVS simulator operate as a cloud-native workload while keeping model management, execution, and data integration independently deployable.
    \item A publishing and execution pipeline for simulation models in which repository CI validates the publishing manifest, the engine loader applies syntactic, structural, and static semantic (well-formedness) validation before every run, and publishing produces immutable versioned artefacts from which an earlier validated model version is selected by pinning its identifier.
    \item An initial case study on the Riga bus route~22 corridor that exercises the lifecycle on an offline simulation model, quantifies single-container engine throughput and concurrent job execution on one Kubernetes node, and varies a multi-modal scenario through manifest and data changes without engine-code changes.
\end{enumerate}

The four contributions together answer the three questions. Each ingredient has a precedent: DEVSML~3.0~\cite{Mittal2017} already grounds a declarative model language in formal DEVS semantics, CI pipelines for simulation software validate declarative metadata~\cite{Chand2025,Reiterer2023}, and immutable versioned artefacts with pinning-based reversion are standard in DevOps and MLOps tooling~\cite{Kreuzberger2023,Karamitsos2020}. The novelty of this paper is their integration within a single model lifecycle; the surveyed platforms each address a subset of them. In the evaluated system the loader's checks run at job start. The Discussion answers each question and states what the evidence supports and what remains untested. The work extends our prior research on DEVS-based simulation engines for digital twins~\cite{Lektauers2025} to their operational lifecycle in the cloud.

\section{Background and Related Work}\label{sec:background}

Running DEVS models as orchestrated cloud workloads brings together two research lines. The DEVS formalism and its simulation engines supply the modelling and execution foundation, and cloud-native simulation and DevOps for models supply the deployment and lifecycle practice.

Table~\ref{tab:related_works_comparison} summarises where existing solutions stand on the dimensions our architecture addresses. The ratings are based on the cited publications: full support means that the publication describes the capability as implemented, partial that it is covered in part or under restrictions, and none that the publication does not report it, which is weaker than a confirmed absence in the tool. ``Declarative spec.'' requires that the executable model itself is specified declaratively, ``Static semantic valid.'' denotes well-formedness checking against the simulation formalism before execution (Section~\ref{subsec:validation}) and does not cover behavioural validation, ``Versioning/rollback'' requires versioning of model artefacts (deployment parameters and container images alone do not count), and ``Model CI/CD'' requires validation of model content before publication. For the proposed approach, ``Model CI/CD'' is partial because its repository CI validates the manifest only, ``Versioning/rollback'' records version storage and selection without an evaluated rollback trace, ``Cloud-native'' records Kubernetes job execution without measured multi-node scalability, and ``Digital-twin focus'' records the intended application without an operational bidirectional twin.

\begin{table}[h]
\centering
\caption{Comparison of related works across the dimensions addressed by the proposed architecture (\fullsupp{}~full support, \partsupp{}~partial, \nosupp{}~none; criteria in the text).}
\label{tab:related_works_comparison}
\footnotesize
\setlength{\tabcolsep}{2pt}
\begin{tabular}{l*{8}{c}}
\toprule
\textbf{Solution}
  & \makecell{Simulation\\support}
  & \makecell{Cloud-\\native}
  & \makecell{Model\\CI/CD}
  & \makecell{Declarative\\spec.}
  & \makecell{Formal\\semantics}
  & \makecell{Static\\semantic\\valid.}
  & \makecell{Versioning/\\rollback}
  & \makecell{Digital-twin\\focus} \\
\midrule
CloudSim \cite{Hewage2024,Andreoli2025} & \fullsupp & \partsupp & \nosupp & \nosupp & \nosupp & \nosupp & \nosupp & \nosupp \\
PerfSim \cite{Khan2023}      & \fullsupp & \partsupp & \nosupp & \partsupp & \nosupp & \nosupp & \nosupp & \nosupp \\
Stoja et al.\ \cite{Stoja2026}   & \fullsupp & \fullsupp & \nosupp & \nosupp & \nosupp & \nosupp & \nosupp & \nosupp \\
Modelica DT \cite{Kumar2022}   & \fullsupp & \fullsupp & \nosupp & \fullsupp & \fullsupp & \nosupp & \nosupp & \fullsupp \\
xDEVS \cite{RiscoMartin2023,RiscoMartin2022}   & \fullsupp & \partsupp & \nosupp & \nosupp & \fullsupp & \nosupp & \nosupp & \nosupp \\
DEVSML 3.0 \cite{Mittal2017}   & \fullsupp & \fullsupp & \nosupp & \fullsupp & \fullsupp & \partsupp & \nosupp & \nosupp \\
KTWIN \cite{Wermann2025}        & \nosupp & \fullsupp & \nosupp & \fullsupp & \nosupp & \nosupp & \nosupp & \fullsupp \\
\midrule
\makecell[l]{\textbf{Proposed} \\ \textbf{Approach} } (\S\ref{sec:model}--\ref{sec:devops}) & \fullsupp & \fullsupp & \partsupp & \fullsupp & \fullsupp & \fullsupp & \fullsupp & \fullsupp \\
\bottomrule
\end{tabular}
\end{table}

\subsection{Digital Twins and DEVS Simulation}

The digital twin concept, introduced by Grieves~\cite{Grieves2017}, describes a virtual representation of a physical system with bidirectional data flow between the physical and digital spaces. Tao et al.~\cite{Tao2019} formalise a five-dimensional model of physical entity, virtual entity, services, data, and connections, and the survey by Fuller et al.~\cite{Fuller2020} maps the enabling technologies.

DEVS~\cite{Zeigler2018} defines atomic models with states, input/output ports, and transition functions, together with coupled models that compose atomic models via port couplings. multiPDEVS~\cite{Foures2018} extends Parallel DEVS to multicomponent systems with nonmodular decomposition and efficient handling of simultaneous events. Mature engines surveyed by Wainer and Govind~\cite{Wainer2024} include ADEVS~\cite{Nutaro2011}, CD++~\cite{Wainer2009}, VLE~\cite{Quesnel2009}, PyPDEVS~\cite{VanTendeloo2015}, and xDEVS~\cite{RiscoMartin2023}, complemented by graphical authoring (DEVS-Suite~\cite{Kim2009}) and DEVS-based co-simulation wrappers (MECSYCO~\cite{Camus2018}). The suitability of DEVS for digital twins has been recognised in the context of digital quadruplets~\cite{Niyonkuru2021}, and Vanommeslaeghe et al.~\cite{Vanommeslaeghe2024} demonstrate co-simulation with Functional Mock-up Interface (FMI). Our work builds on a multiPDEVS engine~\cite{Lektauers2025} that supports hierarchical composition with Parallel DEVS semantics.

\subsection{Cloud-Native Simulation and DevOps for Models}
Simulation engines have shifted from monolithic designs toward modular~\cite{Calheiros2011}, containerised, and orchestrated services~\cite{Stoja2026}. Simulation services are increasingly exposed as microservices~\cite{Hewage2024,Andreoli2025} and deployed through orchestrators such as Kubernetes~\cite{Khan2023,Wermann2025}, which improves scalability and manageability for digital twin and cyber-physical workloads. Current digital twin platforms adopt stateless, cloud-native, service-based architectures~\cite{Kumar2022}. Such frameworks support scalable data integration and distributed execution, but they concentrate on deployment logistics and leave model management aside (Table~\ref{tab:related_works_comparison}).

Closing the gap between deployment and model management requires automated model lifecycles. In software development, process automation is now routine practice~\cite{Kreuzberger2023}, yet model management still lags behind~\cite{daGiao2024}.
CI/CD practices in software and AI development provide versioning, staged validation, delivery, monitoring, and rollback, and DevOps and Machine Learning Operations (MLOps) tooling is gradually extending these capabilities to models~\cite{Kreuzberger2023,Karamitsos2020}.
For simulation software, prior work on GitLab-based CI/CD for large-scale simulation software~\cite{Chand2025} and DevOps models for metadata-driven tools~\cite{Capizzi2020} shows that simulation code and metadata can be integrated into automated pipelines~\cite{Reiterer2023,Colantoni2020}.
Simulation models themselves, however, are not yet consistently treated as first-class DevOps artefacts and still lack widely adopted approaches for declarative specification, structural and static semantic validation, and CI/CD across their cloud lifecycles. Recent work on DevOps, model-driven engineering (MDE), and MLOps documents the same shortfall~\cite{Colantoni2020,Reiterer2023,Subramanya2022}. Complementary work on digital twin programming with semantically lifted states~\cite{Kamburjan2021} ties runtime data to ontologies and formal specifications, reinforcing the case for treating twin models as structured, queryable artefacts. The adoption of SysML~v2~\cite{OMG2025sysml} provides a standard systems-modelling language for such artefacts; its relation to the declarative language proposed here is discussed in Section~\ref{subsec:smdl}.

\section{Model-Centric Representation and Lifecycle}\label{sec:model}

A central design principle of our architecture is treating simulation models as version-controlled artefacts with a well-defined lifecycle. Two declarative languages carry this lifecycle. The Simulation Model Definition Language (SMDL) defines what a model is and how it behaves. The Simulation Scenario Definition Language (SSDL) defines under which configurations a model is studied: it binds registered model versions to parameterisations, datasets, replications, and random seeds. This section covers SMDL, its mapping to the DEVS formalism, and the validation and versioning strategy; SSDL is described together with scenario assembly in Section~\ref{subsec:scenario-assembly}.

\subsection{YAML-Based Model Definition Language}\label{subsec:smdl}

\begin{figure}[t]
    \centering
    \includegraphics[width=0.8\columnwidth]{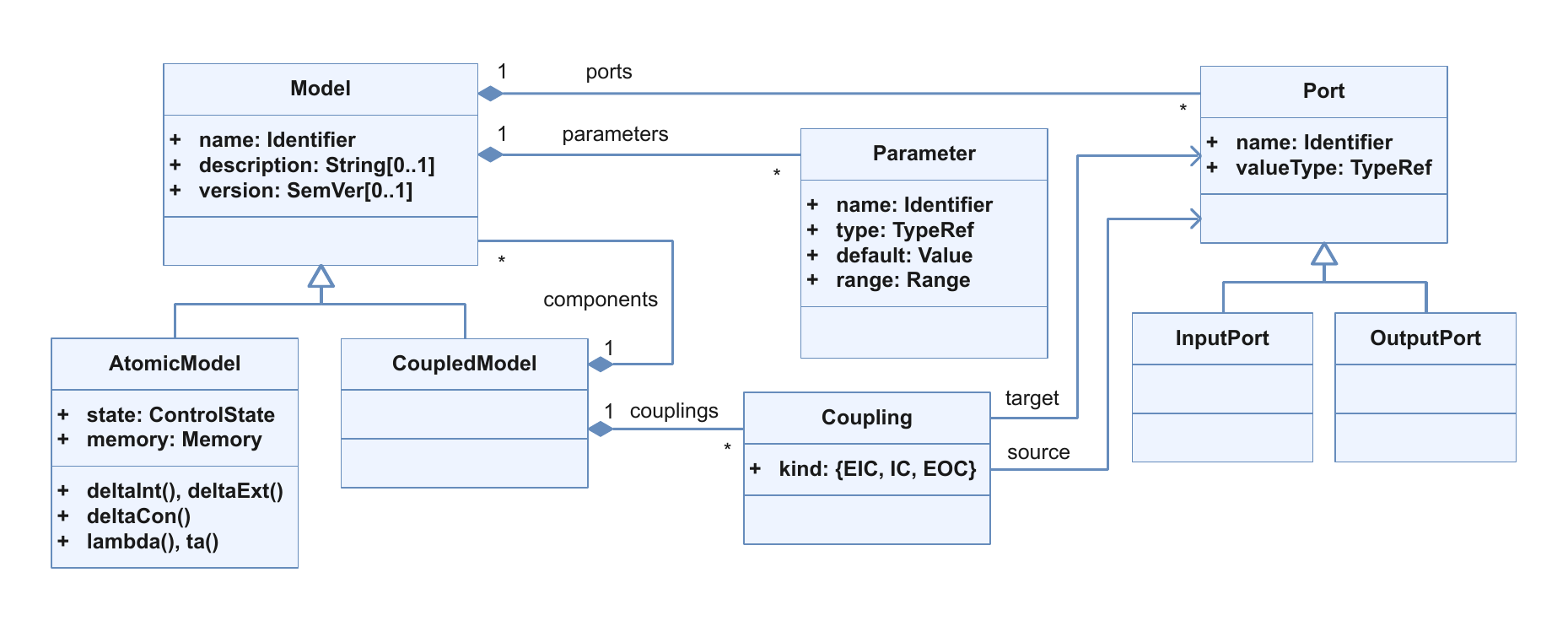}
    \caption{SMDL abstract syntax metamodel (conceptual core). The loader accepts instances of the concrete schema, which realises the subset of Section~\ref{subsec:smdl}. The SMDL Processor transforms conforming documents into multiPDEVS tuples (Eqs.~\ref{eq:multipdevs}--\ref{eq:component}); schema conformance is the structural validation layer, and the well-formedness checks of the mapping form the static semantic layer.}
    \label{fig:smdl-metamodel}
\end{figure}

SMDL specifies DEVS simulation models in YAML, a human-readable, version-controllable format. The language draws on the declarative style of Microsoft's Digital Twins Definition Language (DTDL)~\cite{Microsoft2026} and shares the MDE stance of DEVSML~3.0~\cite{Mittal2017}; YAML files yield line-oriented Git diffs and can be checked in CI with the tooling already used for container manifests. Figure~\ref{fig:smdl-metamodel} shows the abstract syntax: atomic models with control state, memory, and the DEVS characteristic functions, recursively composed coupled models, typed ports, couplings, and parameters. The diagram mirrors the engine's model classes; the concrete schema described next realises a subset of it, and the structural validation layer checks conformance to that schema. The coupling type determines its endpoints: an output port of a component to an input port of a sibling (IC), an input port of the coupled model to an input port of a component (EIC), or an output port of a component to an output port of the coupled model (EOC). The language covers fully declarative models (behaviour configured through parameters of pre-built classes) and hybrid models (behaviour implemented in code and referenced by name).

A model document lists tagged model definitions under \texttt{models:}; the tag selects the modelling style: \texttt{!<discrete\_event>} for process-oriented models, \texttt{!<agent\_based>} for agent populations, and \texttt{!<hybrid>} for agents that flow through a network of blocks. Listing~\ref{lst:smdl-model} is an excerpt of a discrete-event model of a signalised intersection in Riga, reused in the validation walk-through of Table~\ref{tab:validation-walkthrough}. Each tagged entry under \texttt{blocks:} declares one atomic component: the tag (\texttt{source}, \texttt{processor}, \texttt{queue}, \texttt{gate}, \texttt{sink}, and others) selects a pre-built class, the fields configure it, and each \texttt{to:} reference couples the output of the declaring block to the input of the named block. The block syntax leaves ports implicit and expresses successor relationships through \texttt{to:} references, and composition across model documents is not part of the schema; explicit ports and recursive composition appear only in the models that the domain builder of the case study generates, as described below.

\begin{lstlisting}[language=yaml,float=tb,caption={Excerpt of an SMDL discrete-event model of a signalised intersection (first approach only; omissions are marked by comments).},label={lst:smdl-model},basicstyle=\footnotesize\ttfamily,frame=single,breaklines=true,showstringspaces=false]
name: slokas_kurzemes
version: "1.0.0"
time_unit: m
models:
  - !<discrete_event>
    name: intersection
    variables:
      - !<scalar> { name: r1_straight1, value: 22.89 }
      # ... further segment lengths and the signal schedule
    entities:
      - name: car
        attributes:
          - !<triangular> { name: velocity, min: 583.33,
                            mode: 833.33, max: 833.33 }
    blocks:
      - !<source> { name: ar1, entity: car,
                    rate: !<exponential> { mean: 0.9091 },
                    to: r1_approach }
      - !<processor> { name: r1_approach,
                       delay: "${r1_straight1} / ${velocity}",
                       to: qr1 }
      - !<queue> { name: qr1, to: gate_r1 }
      # ... gate on the signal phase, crossing, exit segment
      - !<sink> { name: out_r1 }
\end{lstlisting}

Models whose structure follows from data are specified through a domain manifest, which takes the place of the SMDL model document in an SSDL project. Listing~\ref{lst:yaml-model} shows the top-level structure of the transport manifest of the case study: it declares data sources, simulation parameters, the iterative replanning loop, and the scoring function without referencing engine-internal classes, and the engine's \emph{NetworkBuilder} generates the atomic components and couplings of the DEVS hierarchy from the referenced network data. The manifest therefore contains no component or coupling declarations; it passes through the syntactic and structural layers and the parameter checks of the static semantic layer, whereas the coupling checks described below apply to SMDL model documents such as Listing~\ref{lst:smdl-model}. The listing abbreviates the transit file paths, and the baseline runs of the case study use the manifest without the \texttt{iteration} and \texttt{scoring} blocks.

\begin{lstlisting}[language=yaml,float=tb,caption={Top-level structure of the Riga bus route~22 transport manifest (transit file paths abbreviated).},label={lst:yaml-model},basicstyle=\footnotesize\ttfamily,frame=single,breaklines=true,showstringspaces=false]
network:   { source: geoparquet,
             geoparquet: { path: segments.parquet },
             nodes: { path: nodes.parquet } }
transit:   { lines: { path: ... }, schedules: { path: ... },
             stops: { path: ... } }
demand:    { path: demand.parquet }

simulation:
  horizon:  { value: 24, unit: h }
  randomSeed: 42
  network:  { defaultSpeedLimitKph: 50.0,
              saturationFlowVehPerHourPerLane: 1800.0 }
  transit:  { defaultDwellTime: { value: 30, unit: s } }

iteration:
  enabled: true
  maxIterations: 50
  convergenceWindow: 5
  convergenceMeanTolerance: 0.02
  weightUpdateMethod: MSA
  replanning: { routeProbability: 0.30, modeProbability: 0.10,
                departureTimeShiftProbability: 0.20 }

scoring:
  marginalUtilityOfTravelTimePerHour: -6.0
  modeConstants: { car: 0.0, transit: -0.5 }

output:    { directory: results, formats: [parquet, csv] }
\end{lstlisting}

\subsubsection{Relation to SysML}
With the adoption of SysML~v2 and its kernel language KerML by the Object Management Group~\cite{OMG2025sysml}, a standardised language exists whose constructs can express the structural content of an SMDL document: coupled models correspond to part definitions with nested part usages, ports to port definitions, couplings to connections, and atomic models with few control states to state definitions with transitions. A multiPDEVS domain library in SysML~v2 is therefore feasible, as DEVSys showed for SysML~v1~\cite{Kapos2014}, and we regard it as an alternative concrete syntax for the same semantic domain. Three considerations led us to keep YAML as the artefact that the pipeline validates and stores. First, SysML~v2 does not define time advance, confluent transitions, or multiPDEVS influencer sets; a domain library would encode them as metadata, and the transformation to Eqs.~(\ref{eq:multipdevs})--(\ref{eq:component}) with its well-formedness checks would still be required. Second, the checks described below need only a YAML parser and the engine's loader, which a CI runner can invoke; evaluating SysML~v2 constraints there requires a SysML~v2 implementation and the DEVS domain mapping. Third, the model authors in our setting are transport analysts, for whom we expect a compact YAML syntax to present a lower entry barrier; this expectation has not been evaluated. The natural integration is a SysML~v2 front end or KerML domain library whose instances are exported to SMDL through the standard API, so that models authored in the standard reuse the same pipeline; Reiterer et al.~\cite{Reiterer2023} extend SysML~v2 with DevOps metadata in this spirit.

\subsection{Mapping to DEVS}\label{subsec:mapping}

The \emph{semantic domain} of SMDL is the multiPDEVS formalism of Eqs.~(\ref{eq:multipdevs})--(\ref{eq:component}), and the SMDL Processor realises a model-to-model transformation from the abstract syntax (Figure~\ref{fig:smdl-metamodel}) into it. Structural validation (Section~\ref{subsec:validation}) checks conformance to the implemented SMDL schema described above; static semantic validation then checks the well-formedness conditions listed at the end of this subsection on the transformed instance. These conditions do not establish that the code of a referenced component class conforms behaviourally to DEVS execution semantics, which belongs to the behavioural layer of the validation strategy.

\begin{figure}[htbp]
    \centering
    \includegraphics[width=\columnwidth]{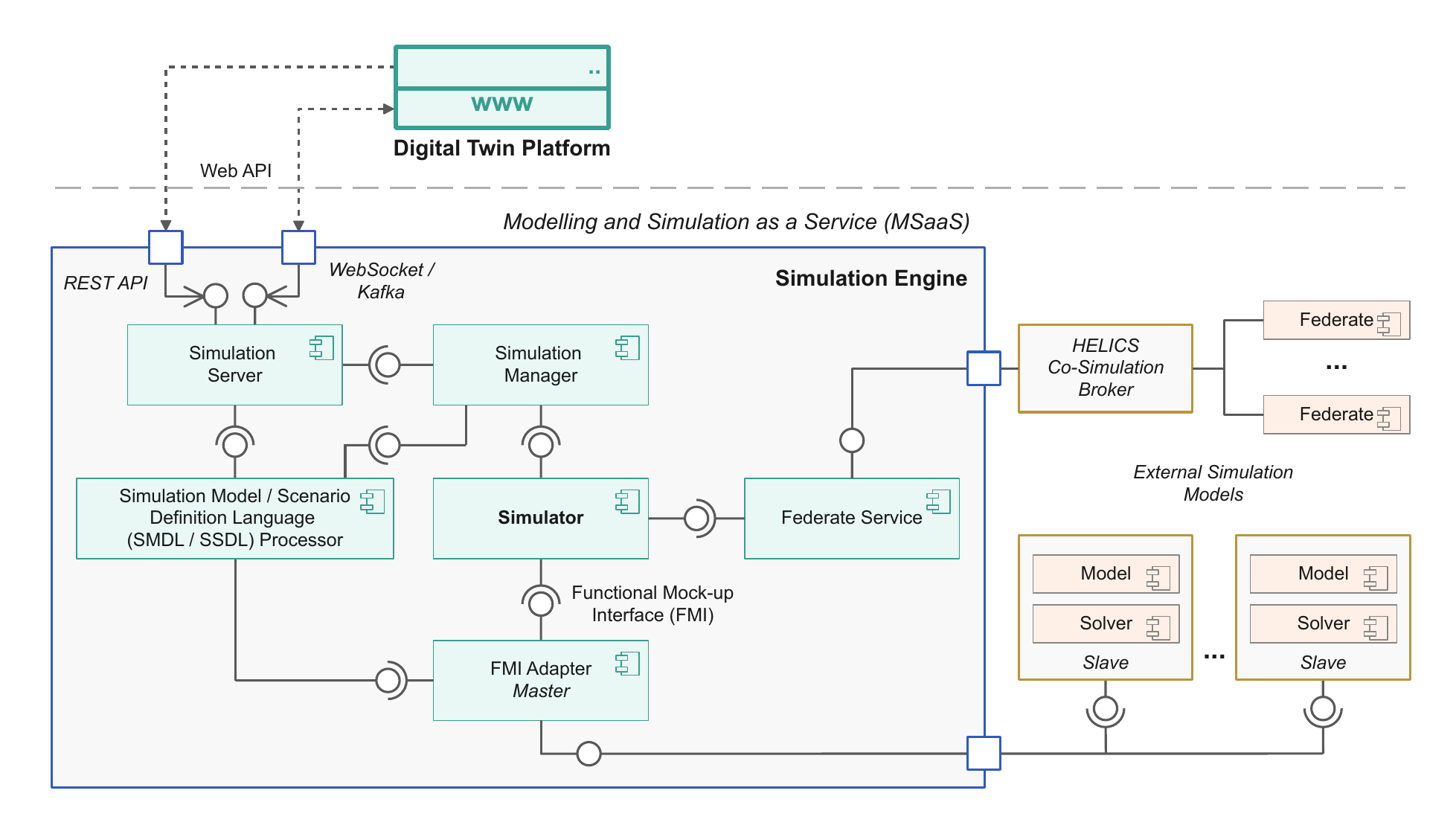}
    \caption{Simulation engine MSaaS architecture with REST, WebSocket and Kafka APIs, SMDL processing, and FMI co-simulation. Adapted, with permission, from~\cite{Lektauers2025}, \textcopyright~2025 IEEE.}
    \label{fig:msaas}
\end{figure}

The SMDL Processor translates a YAML model definition into the internal multiPDEVS~\cite{Foures2018} representation used by the simulation engine, following the Modelling and Simulation as a Service (MSaaS) paradigm~\cite{Lektauers2025}. Figure~\ref{fig:msaas} shows the engine architecture: the SMDL Processor parses the model specification and instantiates DEVS elements, the Simulation Manager coordinates execution following the multiPDEVS abstract simulator of~\cite{Foures2018} as implemented in~\cite{Lektauers2025}, and an FMI adapter supports co-simulation with external Functional Mock-up Units~\cite{Blochwitz2012,Vanommeslaeghe2024}. A Hierarchical Engine for Large-scale Infrastructure Co-Simulation (HELICS) broker~\cite{Hardy2024} further connects the engine to multi-domain federates for scalable cyber-physical co-simulation scenarios.

We formalise the mapping target using the multiPDEVS formalism of~\cite{Foures2018}, with X-Machines~\cite{Holcombe1988} providing the component memory structure as adopted in the engine design of~\cite{Lektauers2025}. A multiPDEVS model is a tuple
\begin{equation}\label{eq:multipdevs}
    \mathrm{multiPDEVS} = (X,\, Y,\, D,\, \{M_d \mid d \in D\}),
\end{equation}
where $X$ and $Y$ are the sets of input and output events and $D$ is the set of component references. Each atomic component $M_d$ is defined as
\begin{equation}\label{eq:component}
    M_d = (\mathbb{S}_d,\, \mathbb{M}_d,\, I_d,\, E_d,\, \delta_{ext,d},\, \delta_{int,d},\, \delta_{con,d},\, \delta_{reac,d},\, \lambda_d,\, ta_d),
\end{equation}
where $\mathbb{S}_d$ is a finite set of control states, $\mathbb{M}_d$ is an explicit memory structure, $I_d \subseteq D$ is the set of influencing components, and $E_d \subseteq D$ is the set of influenced components. The state of a component is the pair $(s,\, m)$ of a control state $s \in \mathbb{S}_d$ and a memory value $m \in \mathbb{M}_d$, and its total state is $((s,\, m),\, e)$, where $e$ is the time elapsed since the last transition and is bounded by the time advance. The time advance $ta_d$, the output function $\lambda_d$, and the internal, external, and confluent transition functions $\delta_{int,d}$, $\delta_{ext,d}$, and $\delta_{con,d}$ operate over the total states of the influencing components $I_d$. The transition functions propose state changes for the influenced components $E_d$, and each influenced component resolves the proposals it receives with its reaction function $\delta_{reac,d}$. Separating the control state $\mathbb{S}_d$ from the memory $\mathbb{M}_d$ admits compact state-machine representations for components with few control states but rich local context, such as agents that carry attributes, history buffers, or physical variables~\cite{Lektauers2025}. The function signatures and the abstract simulator are given in~\cite{Foures2018}, and the extension with control states and memory in~\cite{Lektauers2025}.

The engine executes two kinds of models. The components of a multiPDEVS model interact through the influence relations $I_d$ and $E_d$: a transition of component $d$ reads the states of its influencers and proposes state changes to its influencees, without ports. Classic Parallel DEVS atomic and coupled models with explicit ports~\cite{Zeigler2018} interact through events on port couplings and run on the engine's port-based simulator; in terms of Eq.~(\ref{eq:multipdevs}), such an atomic model corresponds to a multiPDEVS model with a single self-influencing component, $D = \{d\}$ and $I_d = E_d = \{d\}$, whose event sets $X$ and $Y$ are its port events. This correspondence places the port-based models in the formal picture, and the engine performs no such translation. The transport model of the case study consists of port-based models generated by a domain builder; it is not written in SMDL, so the mapping below does not apply to it.

The SMDL Processor maps a \texttt{!<discrete\_event>} document to one flat multiPDEVS model of the form of Eq.~(\ref{eq:multipdevs}), without wrapping the block network into port-coupled models: $D$ is the set of block names, the entities declared under \texttt{entities:} are the items that components hand to each other, and $X$ and $Y$ are empty, because a model document has no external ports. Each block becomes a component tuple of the form of Eq.~(\ref{eq:component}). For fully declarative components, the block tag selects a pre-built class that supplies the control states $\mathbb{S}_d$ and the transition, output, and time-advance functions, and the fields of the block parameterise them; for hybrid components, a script reference supplies the memory $\mathbb{M}_d$ and the code of the functions that operate on it. The sets $I_d$ and $E_d$ are derived from the wiring and never written by the modeller: in this mapping every block influences itself, $d \in I_d \cap E_d$, because its internal transition changes its own state, and a reference \texttt{to:~b} in block $a$ adds $b$ to $E_a$ and $a$ to $I_b$. The handover of an entity from $a$ to $b$ is the proposal of Eq.~(\ref{eq:component}): during its transition, $a$ offers the entity to $b$, which accepts or refuses it according to its own state. For the first two blocks of Listing~\ref{lst:smdl-model}, $I_{\mathtt{ar1}} = \{\mathtt{ar1}\}$, $E_{\mathtt{ar1}} = \{\mathtt{ar1}, \mathtt{r1\_approach}\}$, $I_{\mathtt{r1\_approach}} = \{\mathtt{ar1}, \mathtt{r1\_approach}\}$, and $E_{\mathtt{r1\_approach}} = \{\mathtt{r1\_approach}, \mathtt{qr1}\}$; the time advance of \texttt{ar1} is the next sample of its inter-arrival distribution, and its internal transition creates a \texttt{car} entity and offers it to \texttt{r1\_approach}, whose time advance while it holds an entity is the value of its \texttt{delay} expression, after which it offers the entity to \texttt{qr1}.

Three well-formedness checks read directly from the tuple components and form the core of the static semantic layer: (i)~reference integrity: every \texttt{to:} reference, and every route target of a routing block, must name a block declared in the same model, so that $E_d \subseteq D$ for every component; (ii)~closure of the wiring: every block other than a sink must name at least one receiver, and a sink names none; (iii)~block names must be unique, so that $D$ is a set. Checks~(i) and~(iii) hold for every multiPDEVS instance. Check~(ii) is an SMDL wiring constraint: the formalism admits $E_d = \{d\}$, and the constraint exists because a document has no external interface, so an entity leaving a block without a receiver would be lost silently. The checks are necessary conditions and establish no further property of the formalism; the loader does not check, for instance, for cycles of blocks with zero time advance. In Listing~\ref{lst:smdl-model}, renaming the target of \texttt{ar1} to an undeclared block violates check~(i) and produces the third diagnostic of Table~\ref{tab:validation-walkthrough}.

\subsection{Validation and Versioning}\label{subsec:validation}

Each model definition is versioned using semantic versioning (major.minor.patch): patch for parameter adjustments that do not alter model structure, minor for backward-compatible additions of new components or ports, and major for interface-breaking structural changes. The convention is applied by model authors; the pipeline does not yet check a version increment against the structural change it accompanies (Table~\ref{tab:validation-extensions}). The Model Management Layer maintains a full history of deployed model versions, so that an execution request names the exact model version it uses. A pinned model version fixes one input of a run; the identity of a run also comprises the SSDL document, the input datasets, the random seeds, and the engine build. The SSDL document records the model and dataset references, the replication settings, and the seed. Datasets are referenced by version-scoped object-store paths, immutable by convention, and the engine image by a tag in the job manifest; a tag can move to another image, so a reproducible record needs the image digest, which the platform does not store yet. The artefact bundle of the case study (Data availability statement) records the engine commit, the image digest, the SHA-256 checksum of every input file, and the seeds of every reported run. Such a record reconstructs the inputs of a run exactly; equality of results across Java Virtual Machine (JVM) versions, hardware, and concurrent scheduling was not tested.

Validation is organised in layers whose names we fix here, because the terms are used differently across the MDE and digital-twin communities. \emph{Syntactic validation} checks that the document is well-formed YAML. \emph{Structural validation} checks conformance to the SMDL schema derived from the metamodel of Figure~\ref{fig:smdl-metamodel} (required fields, field types, and admissible tags). \emph{Static semantic validation} checks the well-formedness conditions of the previous subsection on the mapped instance, the availability of referenced entities, resources, and functions, and parameter range constraints, which depend on the model class. All three layers are static: they are decided on the artefact when it is loaded, before any simulation is executed, so they can run as CI stages. In the evaluated implementation the engine's loader performs them in this order, through its YAML parser, the typed deserialisation of the document into the schema classes, and a normalisation pass that checks wiring and parameters; the loader is part of the engine executable, and a separate validation-only tool is not provided. \emph{Behavioural validation}, which checks that simulated behaviour matches the intended semantics of the physical system, satisfies domain invariants, or agrees with observed execution traces, is a distinct runtime concern. The pipeline reported here does not implement behavioural validation. Throughout this paper, ``semantic validation'' refers to the static layer. Extensions towards behavioural validation are discussed below.

Table~\ref{tab:validation-walkthrough} walks model revisions through the three static layers by injecting four faults: three into the bus route~22 manifest and the coupling fault into the SMDL model of Listing~\ref{lst:smdl-model}. Each revision was loaded with \texttt{-{}-scenario=}\textit{file} on an SSDL project that references the faulty model. Each fault is rejected by the earliest layer able to detect it, and the process exits with a non-zero status before simulation starts, so a CI stage that runs this command fails for such a revision. In the evaluated system the three layers run inside the loader at the start of every job, so every executed case-study scenario passed them. The repository pipeline of Section~\ref{sec:devops} checks the manifest against the platform schema; a loader stage that blocks registration of a rejected revision is the intended integration and has not been wired in yet.

\begin{table}[t]
\caption{Fault-injection walk-through: the injected fault, the static layer at which the engine's loader rejects the revision before simulation, and a diagnostic excerpt.}
\label{tab:validation-walkthrough}
\centering
\footnotesize
\begin{tabular}{>{\raggedright\arraybackslash}p{0.34\columnwidth}>{\raggedright\arraybackslash}p{0.13\columnwidth}>{\raggedright\arraybackslash}p{0.41\columnwidth}}
\toprule
\textbf{Injected fault} & \textbf{Rejecting layer} & \textbf{Diagnostic} \\
\midrule
Unbalanced brace in the \texttt{simulation:} block of the bus route~22 manifest & Syntactic & \texttt{at simulation.horizon on line 23, column 21: while parsing a flow mapping} \\
\texttt{horizon.value} given as the string \texttt{"24h"} & Structural & \texttt{at simulation.horizon.value on line 22, column 21: Value '24h' is not a valid floating point value} \\
Coupling whose target names a block absent from the model (check~(i) of Section~\ref{subsec:mapping}) & Static semantic & \texttt{SMDL model 'slokas\_kurzemes': block 'ar1': 'to' names unknown block 'r1\_approach\_missing'} \\
\texttt{defaultDwellTime} with \texttt{value: -30} & Static semantic & \texttt{DurationSpec.value must be finite and non-negative} \\
\bottomrule
\end{tabular}
\end{table}

\subsubsection{Validation Scope and Extensions}\label{subsubsec:validation-scope}

Table~\ref{tab:validation-extensions} lists the richer forms of validation expected in digital-twin engineering, the hook the present architecture provides for each, and the extension required. Two of them are direct CI stages over artefacts that the registry already holds and form the planned next increment of the pipeline: a \emph{version-evolution check} that diffs the SMDL structure of consecutive versions against the declared semantic-versioning increment, and a \emph{regression stage} that re-executes a pinned reference scenario against a new model version and compares key outputs with the stored results of the previous version within a tolerance. The remaining items are left to future work and require, depending on the form, an engine conformance suite with known traces, cross-parameter constraints in the SSDL schema, a property language over the event stream, or live observations from the physical system.

\begin{table}[t]
\caption{Richer forms of validation, the hook the current architecture provides for each, and the extension required. Each extension remains future work for the pipeline evaluated in this paper.}
\label{tab:validation-extensions}
\centering
\footnotesize
\begin{tabular}{>{\raggedright\arraybackslash}p{0.22\columnwidth}>{\raggedright\arraybackslash}p{0.32\columnwidth}>{\raggedright\arraybackslash}p{0.35\columnwidth}}
\toprule
\textbf{Validation form} & \textbf{Existing hook} & \textbf{Required extension} \\
\midrule
Behavioural conformance to DEVS execution semantics & Engine implements multiPDEVS~\cite{Lektauers2025} & Conformance suite of reference models with known traces, run as an engine-level CI stage \\
Outputs vs.\ expected properties or domain invariants & Result callback and Kafka event stream (Section~\ref{sec:devops}); post-run invariant report of the transport engine (Section~\ref{subsec:results}) & Pipeline stage that reads the report and fails the job or release; invariants declared in SSDL (e.g.\ traveller conservation, link load $\le$ capacity) \\
Model assumptions vs.\ runtime observations & Data Integration Service; planned live bus-location feed & Drift detection comparing simulated and observed indicators per model version \\
Scenario-level consistency (inconsistent parameter combinations) & SSDL processor resolves parameters at dispatch & Cross-parameter constraints in the SSDL schema, evaluated before job creation \\
Temporal and causal constraints & Timestamped lifecycle and progress events on Kafka & Property specification (e.g.\ temporal logic) checked over the event stream, with simulation-event traces where a property needs them \\
Model-evolution consistency across versions & Semantic-versioning rules; immutable registry history & Structural diff between versions checked against the declared version increment \\
Regression validation after updates & Immutable artefacts; reproducible pinned runs & Golden-output comparison of a pinned scenario between consecutive versions \\
\bottomrule
\end{tabular}
\end{table}

\section{Cloud-Native Simulation Architecture}\label{sec:architecture}

\begin{figure}[t]
    \centering
    \includegraphics[width=\columnwidth]{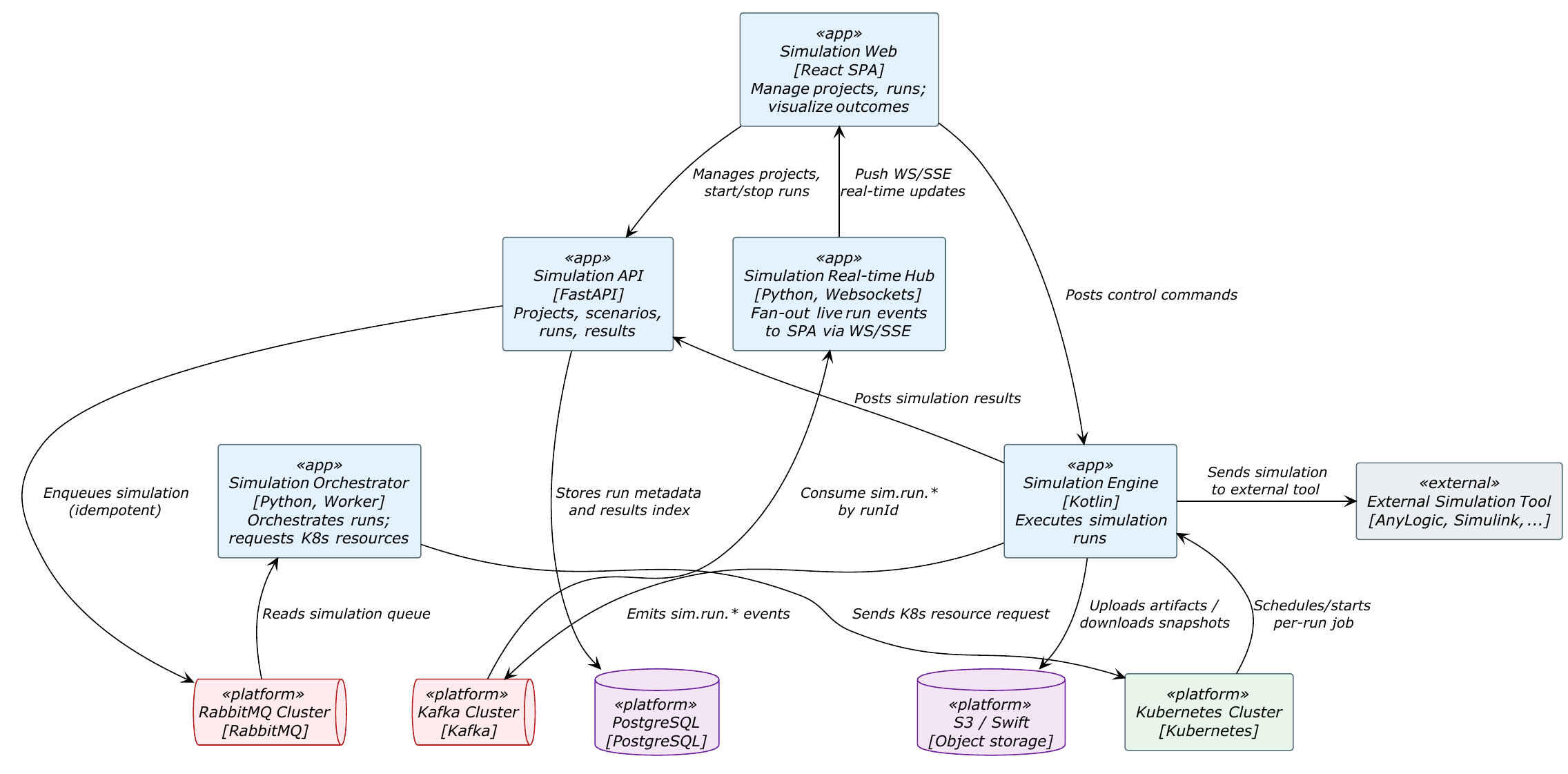}
    \caption{Platform architecture (C4 component view): the microservices and their REST and Kafka interactions. The assignment of services to the four layers is given in the text.}
    \label{fig:platform-architecture}
\end{figure}

The platform is organised into four service layers, summarised as a C4 component view in Figure~\ref{fig:platform-architecture}:

\begin{enumerate}
    \item \textit{Model Management Layer}: storage, versioning, validation, and retrieval of simulation model definitions.
    \item \textit{Simulation Execution Layer}: lifecycle management of simulation runs, including instantiation, execution, and termination of DEVS simulation engine instances.
    \item \textit{Data Integration Layer}: real-time data ingestion from physical systems and external data sources, feeding live data into running simulations.
    \item \textit{Results and Streaming Layer}: collection of simulation outputs and streaming of results to consuming services, dashboards, and analytics pipelines.
\end{enumerate}

Each layer is realised as one or more containerised microservices on Kubernetes that communicate through REST for request--response interactions and through Apache Kafka for events.

The services of the Model Management and Simulation Execution layers are described in the next section. The two remaining layers carry the live data flow. The \textit{Data Integration Service} ingests data from Internet-of-Things (IoT) sensors over the Message Queuing Telemetry Transport (MQTT) protocol, external APIs, databases, and batch files, normalises each event into a DEVS external input, and publishes it to Kafka.

The \textit{Results Streaming Service} delivers simulation outputs over WebSocket, Kafka, InfluxDB, and REST, tagged with run identifiers, model versions, and timestamps for traceability.

\section{DevOps Pipeline for Simulation Models}\label{sec:devops}

Four services realise the DevOps pipeline. A \emph{Simulation Model Registry} and a \emph{Simulation Model Publishing} worker implement the Model Management layer by cataloguing models and ingesting model repositories. A \emph{Simulation API}, which manages projects, scenarios, experiments, and jobs, and a \emph{Simulation Orchestrator}, which materialises each job as a Kubernetes workload, implement the Simulation Execution layer. RabbitMQ delivers each work item (publish, delete, dispatch) to a single consumer, Kafka fans progress and lifecycle events out to multiple listeners, and an S3-compatible object store holds all artefacts.

\subsection{Model Publishing and Registration}

\begin{figure}[h]
    \centering
    \includegraphics[width=1\textwidth]{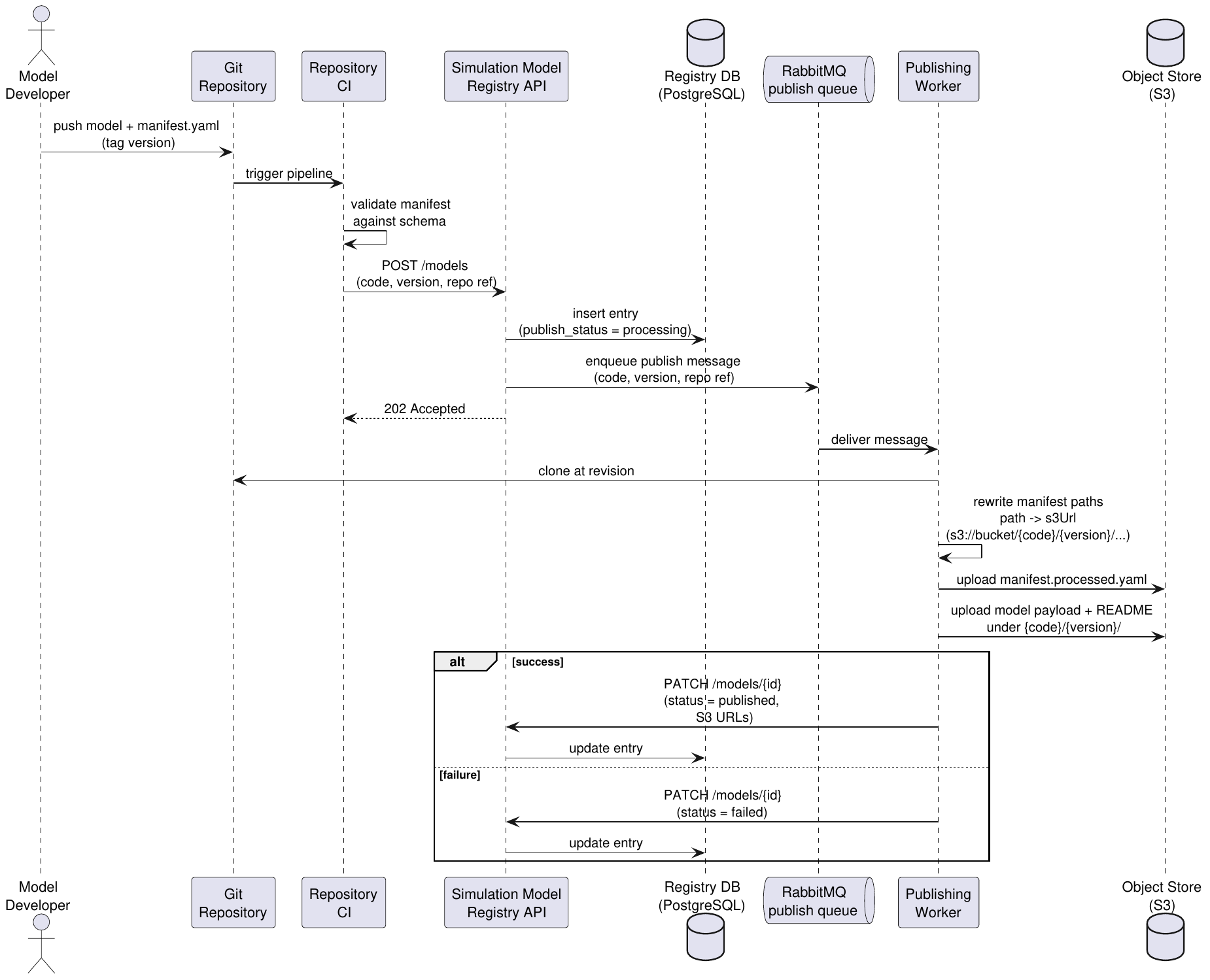}
    \caption{Publishing flow of a simulation model: from Git push and CI validation (manifest schema check) to versioned S3 artefacts and a published registry entry.}
    \label{fig:model-publishing}
\end{figure}

Each simulation model is authored in a dedicated Git repository with a \texttt{manifest.yaml} that describes metadata, component sources, inputs, outputs, parameters, and runtime requirements. A repository-level CI pipeline validates the manifest against the platform schema and registers the model with the registry, which creates a catalogue entry keyed by (\emph{code}, \emph{version}) and queues a publish message. The publishing worker clones the repository at the requested revision and uploads the processed manifest, model payload, and documentation under a version-scoped object-store prefix. The registry record is then marked \texttt{published} with its artefact URLs, or \texttt{failed}, in which case the author corrects the source and publishes a new version. Earlier releases remain immutable in catalogue and object storage, which underpins the reproducibility of downstream experiments. Figure~\ref{fig:model-publishing} summarises this stage. Loader-based model validation is not yet a step of this pipeline; in the evaluated system it runs when a job starts.

\subsection{Scenario Assembly and Job Dispatch}\label{subsec:scenario-assembly}

\begin{figure}[t]
    \centering
    \includegraphics[width=0.85\columnwidth]{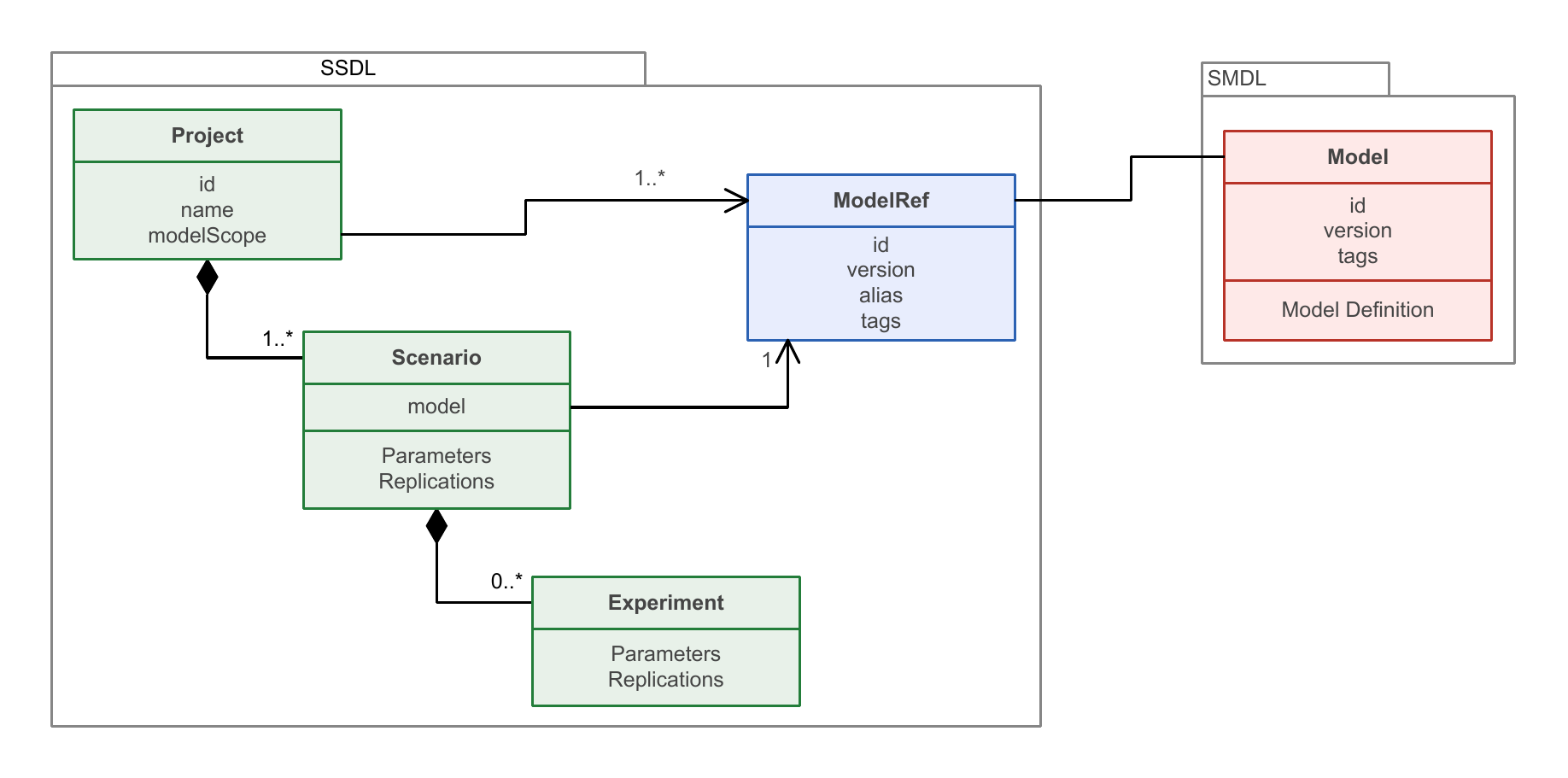}
    \caption{SSDL structure and its binding to SMDL model versions via \emph{ModelRef}.}
    \label{fig:ssdl-structure}
\end{figure}

A published model is consumed through three entities that the Simulation API persists: a \emph{SimulationProject} groups related work, a \emph{Scenario} binds a registered model version to a concrete parameterisation and an optional dataset, and an \emph{Experiment} defines a parameter sweep over its parent scenario. Figure~\ref{fig:ssdl-structure} shows this structure. When execution is triggered for a project, a subset of its scenarios, or individual experiments, the API performs three steps for each selected scenario.

First, it generates an SSDL document that inlines the project metadata, the model locator resolved by the registry, and the scenario configuration with its experiments expanded in place. Second, it uploads the document to a version-scoped object-store path, so that the engine receives a self-contained, versioned input. Third, it creates a \emph{SimulationJob} record and queues a dispatch message with the job identifier and the scenario, model, and optional dataset URLs for a single orchestrator worker.

The API thus decides \emph{what} to run, and the orchestrator decides \emph{where} and \emph{how} to run it. The dispatch message carries only identifiers and immutable object-store references, so a redelivered message describes the same inputs. Delivery to a single consumer does not guarantee exactly-once execution, and the handling of redelivered messages was not exercised in the evaluation.

\subsection{Engine Adaptations for Cloud Deployment}

The Simulation Orchestrator renders a base engine manifest from object storage into a job-specific Kubernetes specification: resource names carry the job identifier, so concurrent runs do not collide, and the job metadata is injected as environment variables. The orchestrator records the in-cluster service address of the engine on the \emph{SimulationJob} record for control operations and monitoring. It supports two storage modes: \emph{S3 mode} passes object-store URLs and credentials to an engine pod that fetches every input at start-up, and \emph{local mode} pre-fetches artefacts to a shared persistent volume where object-store egress is expensive.

The engine shares no database with the Simulation API. On termination it posts the result payload, an object-store URL for larger artefacts, a content type, a byte size, and a SHA-256 checksum to the Simulation API, which attaches the record to the originating experiment. Lifecycle and progress events (engine ready, iteration progress, failure, deletion) are additionally published to Kafka for consumers such as the UI, monitoring, and result archival. A cleanup worker in the orchestrator reconciles Kubernetes state with the Simulation API and deletes the resources of jobs left in the \texttt{ERROR} state beyond a configurable threshold, so that resource reclamation does not depend on a crashed pod. The manifest-driven job rendering, the dual storage modes, the callback-based result protocol, and the reconciliation worker are the adaptations required to operate a classical DEVS engine as a cloud-native workload. They externalise the inputs and results of a run; the simulation state stays in the pod's memory for the duration of the job, a failed job is re-run from its recorded inputs, and checkpoint, restart, or migration of a running simulation are not provided.

\section{Case Study and Evaluation}\label{sec:casestudy}

The Riga bus route~22 corridor is the first instantiation of our planned city-wide multi-modal transport digital twin; it exercises the architecture end-to-end while remaining tractable to analyse in depth. The case study extends our earlier DEVS-based geosimulation framework for public transport analysis and planning~\cite{Lektauers2014} to the cloud-native DevOps setting presented here. The study is offline: its inputs are the published network and schedule, one day of historical e-ticket validations, and a synthetic car demand, with no live connection to the physical system. Table~\ref{tab:claims} maps each claim examined in this section to its evidence and to the boundary of that evidence. The intersection model of Listing~\ref{lst:smdl-model} exercises the SMDL transformation and wiring checks; the transport case study evaluates the shared lifecycle and execution infrastructure with a domain-generated Parallel DEVS model.

\begin{table}[!htb]
\caption{Claims examined in the case study, the evidence presented for each, and the boundary of that evidence.}
\label{tab:claims}
\centering
\footnotesize
\begin{tabular}{>{\raggedright\arraybackslash}p{0.27\columnwidth}>{\raggedright\arraybackslash}p{0.29\columnwidth}>{\raggedright\arraybackslash}p{0.33\columnwidth}}
\toprule
\textbf{Claim} & \textbf{Evidence} & \textbf{Boundary} \\
\midrule
Static validation rejects malformed artefacts before simulation execution (first question) & Four injected faults rejected by the engine's loader (Table~\ref{tab:validation-walkthrough}) & Illustrative inputs; no fault-detection rates over a corpus of revisions; loader not yet a repository CI stage \\
\addlinespace
The lifecycle from registration to result callback is feasible (second question) & Registration, publishing, dispatch, and callback stages exercised on the project's Kubernetes deployment (Section~\ref{subsec:results}) & Functional exercise; stage latencies and failure handling not measured \\
\addlinespace
Scenarios vary without engine-code changes (third question) & Baseline manifest and five further demand levels executed on one engine build (Sections~\ref{subsec:results} and~\ref{subsec:engine-performance}) & Parameter and data changes; structural model revisions across registry versions not traced \\
\addlinespace
Engine throughput permits repeated simulation passes (third question) & Demand sweep with six levels and three runs per level (Section~\ref{subsec:engine-performance}) & Single simulation pass; replanning cost and convergence not measured \\
\addlinespace
Scenario jobs execute concurrently under Kubernetes (second question) & One to six concurrent jobs, five batches per level (Section~\ref{subsec:cluster}) & One node; jobs submitted with \texttt{kubectl}, outside the orchestrator \\
\bottomrule
\end{tabular}
\end{table}

\subsection{Application Context}

Bus route~22 links the centre of Riga (Abrenes street) with Riga International Airport and carries airport passengers and residents along the corridor; the timetable serves 23 stops towards the airport and 17 towards the centre. The trolleybus route with the same public number runs in a different part of the city and is excluded by selecting the route identifier \texttt{riga\_bus\_22} of the timetable feed. The planned digital twin is intended to support transport planners with demand--supply equilibration, transit operations evaluation, and scenario testing for service changes; the case study exercises the architecture on a workload that combines multi-modal routing, iterative agent-based equilibration, and schedule-driven transit operations.

The scenario is configured entirely through the transport manifest shown in Listing~\ref{lst:yaml-model}. Input data is stored in GeoParquet format: a road network of 74{,}508 directed segments and 40{,}746 nodes derived from an OpenStreetMap extract~\cite{OpenStreetMap2026}, and the two line directions of the route with 37 stops and 215 weekday departures from the published General Transit Feed Specification (GTFS) feed~\cite{RigasSatiksme2026a}, in the weekday timetable in force since 20~April 2026. Transit demand is generated from the open e-ticket validation log~\cite{RigasSatiksme2026b} of bus route~22 for Tuesday, 9~June 2026, a day within the validity period of the timetable: each of the 5{,}697 tap-in events becomes one synthetic transit traveller who departs at the validation time. The log records no stop, and its direction field carries the same value for about 97\% of the tap-ins of that day, so the line direction of a traveller is drawn at random in proportion to the scheduled departures per direction, the boarding stop is sampled uniformly along that direction, and the alighting stop is sampled among the downstream stops with a probability that decays exponentially with the number of stops travelled. The traveller count equals the number of recorded validations; the log omits fare-free passengers, so real loading exceeds it by an unknown amount. Car demand is synthesised as 12{,}000 trips with random origin--destination pairs among the 6{,}449 network nodes within a 600~m straight-line walking catchment~\cite{ElGeneidy2014} of a bus route~22 stop that belong to the largest strongly connected component of the road graph; departure times follow the same empirical daily profile as the transit demand. The combined scenario has 17{,}697 traveller demands.

\subsection{DEVS Model Architecture}

\begin{figure}[t]
    \centering
    \includegraphics[width=\columnwidth]{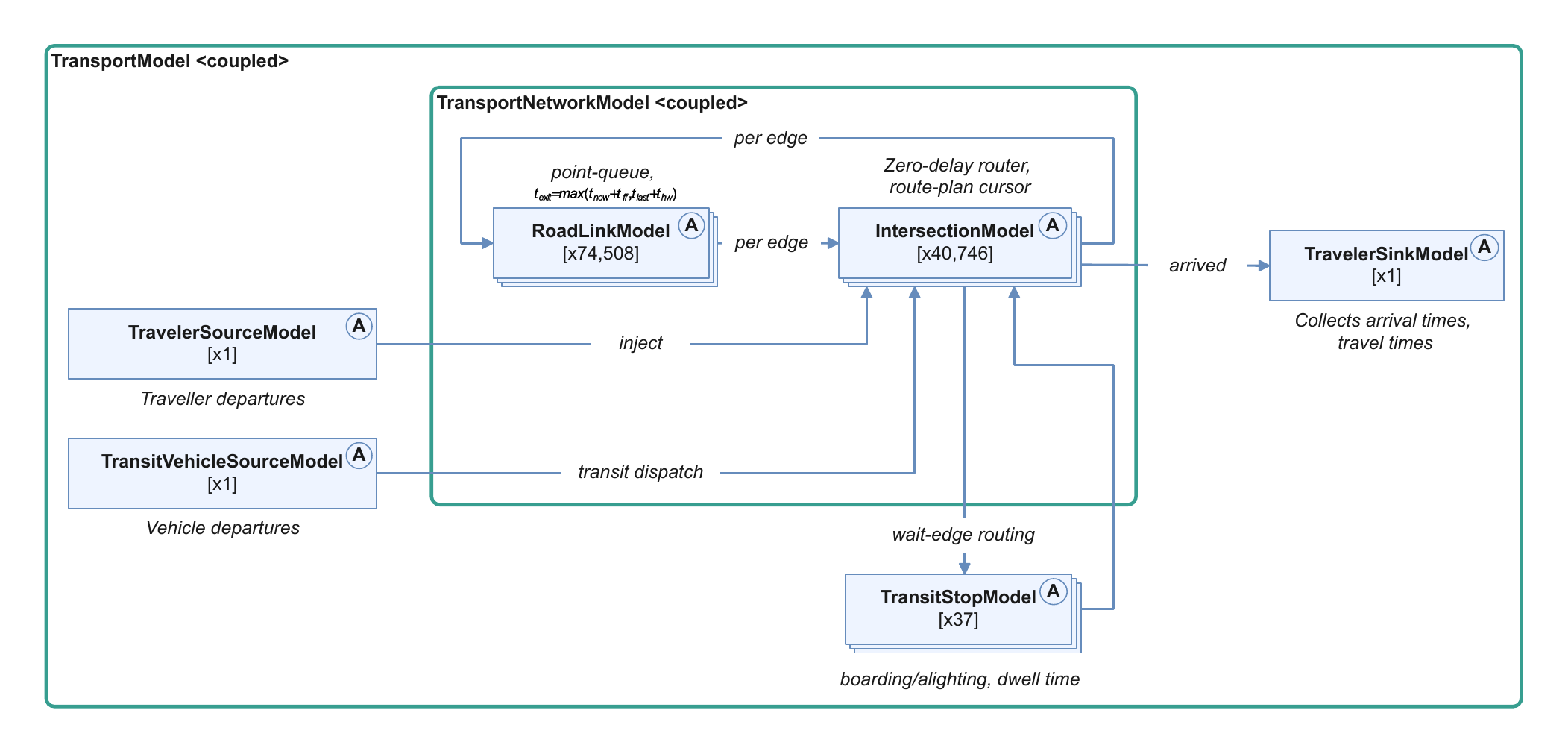}
    \caption{Hierarchical DEVS decomposition of the bus route~22 scenario. The inner \emph{TransportNetworkModel} holds one \emph{RoadLinkModel} per directed segment and one \emph{IntersectionModel} per node; the couplings labelled \emph{per edge} connect each link to the intersection at its end node and each intersection to its outgoing links, one coupling pair per segment. The outer \emph{TransportModel} adds transit-stop, source, and sink models. Arrows are port couplings labelled with the event type; stacked silhouettes denote multiplicity \texttt{[×N]}; the circled \texttt{A} marks atomic models.}
    \label{fig:devs-hierarchy}
\end{figure}

The transport simulation is assembled as a two-level hierarchical coupled DEVS model (Figure~\ref{fig:devs-hierarchy}). The inner coupled model, \emph{TransportNetworkModel}, couples one \emph{RoadLinkModel} per directed segment and one \emph{IntersectionModel} per node; the outer coupled model, \emph{TransportModel}, adds one \emph{TransitStopModel} per stop, the traveller and transit-vehicle sources, and the sink. Six types of atomic model appear in total. The case study uses port-coupled Parallel DEVS atomic models, which the same engine executes; the multicomponent form of multiPDEVS pays off when many components read each other’s state, whereas each link in this sparse FIFO pipeline is influenced by at most one or two upstream components.

\emph{RoadLinkModel} (one per segment) implements mesoscopic point-queue dynamics: each link maintains a vehicle queue ordered by computed exit times, with exit times derived from free-flow travel time, a minimum headway from a saturation flow of 1800~veh/h/lane, and a storage capacity from a 7.5~m jam spacing. \emph{IntersectionModel} (one per node) is a zero-delay router that consults each agent's route plan and forwards the vehicle to the next edge or to an \emph{arrived} port. \emph{TransitStopModel} (one per stop) handles boarding and alighting with a dwell time of $t_{door} + \max(n_b r_b, n_a r_a)$ ($t_{door}=5$~s, $r_b=3$~s/pax, $r_a=2$~s/pax). \emph{TravelerSourceModel} and \emph{TransitVehicleSourceModel} inject travellers and dispatch vehicles on schedule; \emph{TravelerSinkModel} records arrivals. A \emph{NetworkBuilder} constructs the hierarchy automatically from the network graph and integrates transit through synthetic \emph{wait} and \emph{stop} edges.

\subsection{Multi-Modal Routing and Iterative Equilibration}

Car routes are computed by time-dependent Dijkstra on the road graph using the current iteration's edge travel time estimates. Transit routes enumerate candidate (line, boarding stop, alighting stop) triples over the published schedule, scoring each by walking access and egress, in-vehicle travel time, and accumulated dwell times against the same time-dependent edge weights, so that router and simulator remain consistent. Each transit ride is materialised as a synthetic \emph{wait}-edge identifier that routes the traveller to the appropriate \emph{TransitStopModel}.

Demand--supply equilibration follows a MATSim-style~\cite{Horni2016} co-evolutionary loop reimplemented inside the DEVS engine so that network simulation, scoring, and replanning all share the same DEVS event semantics, event stream, and versioned manifest. Each iteration runs the full DEVS simulation, scores plans with a Charypar--Nagel utility~\cite{Charypar2005} (non-arrival penalty $-1000$), stochastically replans a fraction of travellers (30\% route, 10\% mode, 20\% $\pm 15$~min departure shift), and updates edge travel times via the Method of Successive Averages. Convergence is tested on per-iteration score statistics: after each iteration $k$ the mean $\bar{u}_k$ and the sample standard deviation $\sigma_k$ of the scores of all traveller plans are recorded, with a plan that produced no arrival assigned the configured penalty, and the loop stops after at least five iterations when, over the last five iterations, $\max_k \bar{u}_k - \min_k \bar{u}_k \le 0.02$ and $\max_k \sigma_k - \min_k \sigma_k \le 0.02$. Both tolerances are absolute, in the utility units of the scoring function (the manifest sets the first, and the second keeps its default), and the second condition requires the dispersion of scores across travellers to have stabilised; a large but steady dispersion satisfies it. Utility coefficients are listed in the manifest of Listing~\ref{lst:yaml-model}. The measurements reported below use a single simulation pass without replanning, so the convergence behaviour of the loop is not evaluated in this paper.

\subsection{Simulation Results}\label{subsec:results}

\begin{figure}[h]
    \centering
    \includegraphics[width=1\textwidth]{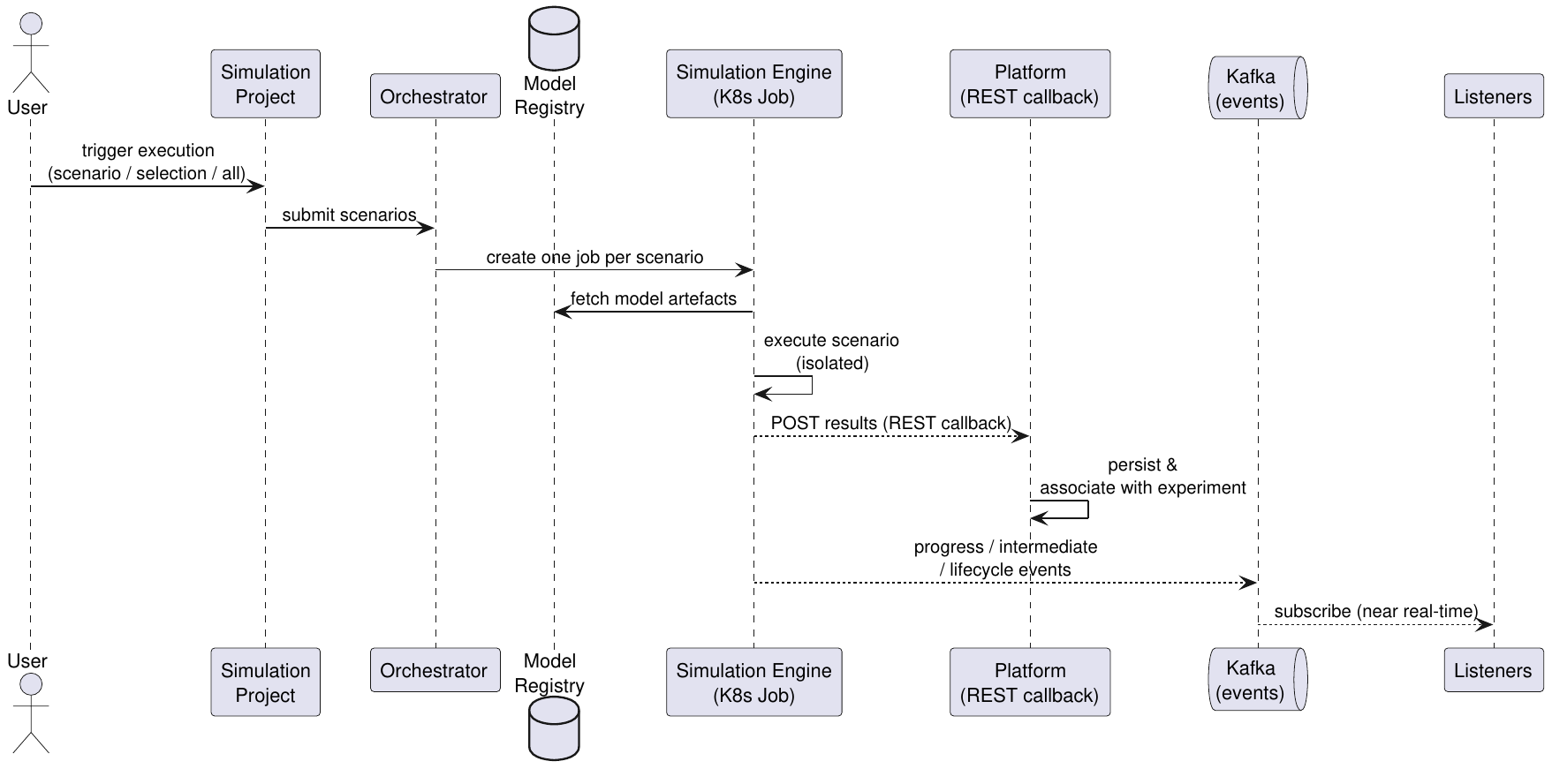}
    \caption{Execution flow of a published model: one Kubernetes job per scenario, results returned by REST callback, progress and lifecycle events streamed over Kafka.}
    \label{fig:execution-flow}
\end{figure}

The published bus route~22 model is executed through the publishing and execution pipeline, which materialises one Kubernetes job per scenario (Figure~\ref{fig:execution-flow}) and so isolates scenarios at the container level. The registration, publishing, dispatch, and callback stages were exercised on the project's Kubernetes deployment, and the measurements that follow were made on a dedicated single-node cluster specified in Section~\ref{subsec:engine-performance}.

Table~\ref{tab:results} summarises the scenario parameters and single-run results for the Riga bus route~22 simulation with a 24-hour horizon, matching the full day of the validation log.

\begin{table}[t]
\caption{Scenario parameters and single-run results for Riga bus route~22 with transit demand derived from e-ticket validations for 9~June 2026; travel times are grouped by requested mode.}
\label{tab:results}
\centering
\begin{tabular}{ll}
\toprule
\textbf{Metric} & \textbf{Value} \\
\midrule
Network segments (DEVS link models) & 74{,}508 \\
Intersection models & 40{,}746 \\
Transit stops / line directions / departures & 37 / 2 / 215 \\
Total DEVS atomic models & 115{,}294 \\
Traveller demand (car / transit) & 17{,}697 (12{,}000 / 5{,}697) \\
Simulation horizon & 24 h \\
\midrule
Travellers arrived (car / transit) & 17{,}676 (11{,}986 / 5{,}690) \\
Median travel time (car) & 4.7 min \\
Median travel time (transit) & 8.6 min \\
P90 travel time (car) & 9.1 min \\
P90 travel time (transit) & 15.4 min \\
Transit stop visits & 4{,}269 \\
Passengers boarded / alighted & 4{,}824 / 4{,}824 \\
\bottomrule
\end{tabular}
\end{table}

Of the 17{,}697 travellers in the demand file, 17{,}676 arrive within the 24-hour horizon: 11{,}986 of 12{,}000 car trips and 5{,}690 of 5{,}697 transit trips (99.9\% of each). The engine's post-run invariant report accounts for the remaining 21, all still under way when the horizon ends (19 on links, one waiting at a stop, one alighting); no traveller is skipped at the source, discarded at an intersection, or rejected on a link, and all twelve checks of the report pass. Of the arriving transit travellers, 4{,}821 ride the bus and 869 complete the trip on foot, because the schedule-aware router returns a walking plan when it reaches the destination earlier than the next departure. Thirteen vehicles find the first link of their path full and wait at its entry until a slot becomes available. The report prompted this correction: on an earlier engine build the scenario failed the traveller-conservation check because five buses leaving a stop into an occupied single-vehicle link were dropped with their 22 passengers; the link model now holds such a vehicle at the entry, and all measurements use the corrected build. The report is an engine diagnostic (Table~\ref{tab:validation-extensions}): the engine writes it with the results and exits with status zero regardless of the report, and no pipeline stage reads it yet, so a completed job does not demonstrate that the domain invariants hold.

\begin{figure}[t]
    \centering
    \includegraphics[width=\columnwidth]{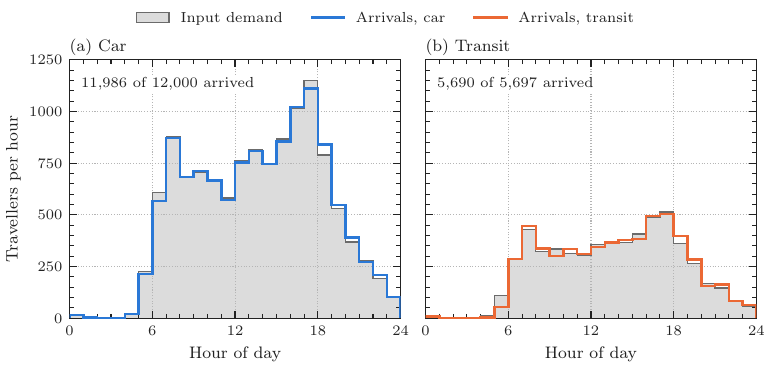}
    \caption{Hourly input demand (grey shading) and engine arrivals (coloured line) for (a)~car and (b)~transit travellers in the baseline run, grouped by the mode requested in the demand file.}
    \label{fig:arrival-profile}
\end{figure}

Figure~\ref{fig:arrival-profile} compares the per-hour input demand against engine arrivals by mode. Both modes share the empirical daily profile recovered from the validation log, with a morning peak (07:00--08:00) and an afternoon peak (16:00--18:00). The hourly arrival profiles broadly follow the prescribed demand profile of each mode. This agreement is a consistency check on the run, since the arrivals are generated from that demand, and does not constitute independent empirical validation of the model.

Car trips are shorter than transit trips (medians of 4.7 and 8.6~min): a car trip connects two catchment nodes directly, and a transit trip includes the wait at the stop. Congestion is sparse. The extract contains many short segments: 53\% of the links are shorter than twice the jam spacing and hold a single vehicle, so such a link reaches the congestion threshold whenever one vehicle is on it at a five-minute sample. Every link that reaches the threshold during the run is of this kind (1{,}017 links, 1.4\% of the network, at most 22 at the same sample), and no link with room for more than one vehicle reaches it. The Discussion returns to this sub-capacity regime.

\subsection{Single-Container Engine Performance}\label{subsec:engine-performance}

Iterative equilibration and interactive scenario exploration require each simulation run to complete in a small fraction of the real-time duration it represents. We ran the 24-hour bus route~22 scenario with the e-ticket-derived transit demand and synthetic car demand at six levels (0--24{,}000 trips, 5{,}697--29{,}697 travellers in total), measuring end-to-end wall-clock runtime of the engine process, including JVM start-up, input parsing, DEVS network construction, execution, and result export. Each level was run three times in a fresh container of the engine image on the cluster node (Intel Core i5-13600K, 20 logical CPUs, 32~GB RAM, Linux Mint~22.3, Docker~29.8, 4~GB JVM heap) without CPU limits. Arrival counts are identical across repetitions because the manifest fixes the random seed, and the standard deviation of runtime is below 1.5~s at every level. These single-container measurements isolate engine cost from cluster scheduling, which the next subsection adds.

\begin{figure}[t]
    \centering
    \includegraphics{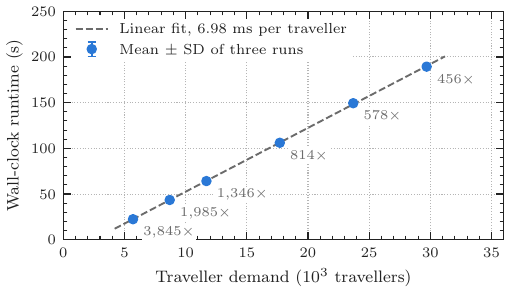}
    \caption{Runtime of the 24-hour bus route~22 simulation vs.\ traveller demand, annotated with the real-time speedup. Points are means of three runs with sample-standard-deviation error bars (0.03--1.4~s, smaller than the markers); the dashed line is the least-squares fit.}
    \label{fig:engine-scaling}
\end{figure}

Figure~\ref{fig:engine-scaling} reports the results. Runtime is approximately linear in the number of car travellers over the tested range, with a least-squares slope of 6.98~ms per additional traveller. The transit-only run of 5{,}697 travellers takes 22.5~s: about 3~s for JVM start-up and input loading, 3~s for the construction of the 115{,}294 atomic models, 12~s for the simulated day, and 4~s for the export of the results. Only the simulation phase grows with demand, reaching 179~s with 24{,}000 cars, while the other phases stay within half a second of their transit-only durations. The baseline scenario of 17{,}697 travellers completes in 106.1~s, a real-time speedup of 814$\times$; the lightest configuration runs at 3{,}845$\times$ real time and the heaviest (29{,}697 travellers, 189.4~s) at 456$\times$. Across the sweep the engine processes 157--254 input travellers per second of wall-clock time, including all framework overhead; this rate divides the size of the demand file by the runtime and so counts travellers who do not arrive. For the iterative replanning loop, the simulation pass of each iteration takes about two minutes at the baseline and three minutes at the heaviest configuration, so the 50 iterations allowed by the manifest of Listing~\ref{lst:yaml-model} would take 1.5 to 2.5~hours of wall-clock time for the simulation passes alone before the cost of replanning is added.

\subsection{Concurrent Execution on a Kubernetes Cluster}\label{subsec:cluster}

We measured the cost of scheduling scenario jobs through Kubernetes and the interference between co-located jobs on a single-node k3s cluster (Kubernetes~v1.36.4, containerd~2.3) on the node of the single-container measurements. The engine image bundles the engine distribution with the published bus route~22 model and its SSDL scenario, so each job is a self-contained Kubernetes \texttt{Job} that runs the 24-hour baseline of Table~\ref{tab:results} with a request of two CPUs and 4.5~GiB of memory, a memory limit of 4.5~GiB, and no CPU limit, so that a job may use idle cores. For each concurrency level $N \in \{1, 2, 4, 6\}$ we submitted $N$ jobs with a single \texttt{kubectl apply} invocation, waited for all of them to finish, and repeated the batch five times, after one discarded warm-up job had unpacked the image on the node; six is the largest level whose memory requests fit the node. Three quantities are recorded from timestamps written by the container entry point: \emph{start-up latency}, from submission to the first instruction of the container; \emph{engine time}, the wall-clock time of the engine process as defined above; and the \emph{makespan} of the batch, from submission until its last job ends. Aggregate throughput is $N \times 17{,}697$ input travellers divided by the makespan.

\begin{table}[t]
\caption{Concurrent execution of the bus route~22 baseline on one Kubernetes node. Start-up latency and engine time are means over all jobs of a level (5, 10, 20, and 30 jobs); makespan and throughput are means over five batches; sample standard deviations of the timing columns in parentheses.}
\label{tab:cluster}
\centering
\setlength{\tabcolsep}{4.5pt}
\begin{tabular}{rcccc}
\toprule
\textbf{Jobs} & \textbf{Start-up latency} & \textbf{Engine time per job} & \textbf{Batch makespan} & \textbf{Throughput} \\
 & (s) & (s) & (s) & (input travellers/s) \\
\midrule
1 & 1.9 (0.5) & 106.2 (1.7) & 108.1 (1.5) & 164 \\
2 & 1.8 (0.5) & 118.3 (0.9) & 120.7 (1.0) & 293 \\
4 & 1.0 (0.2) & 135.6 (2.6) & 138.9 (0.7) & 510 \\
6 & 1.3 (0.6) & 152.8 (4.3) & 158.2 (2.7) & 671 \\
\bottomrule
\end{tabular}
\end{table}

\begin{figure}[t]
    \centering
    \includegraphics[width=\columnwidth]{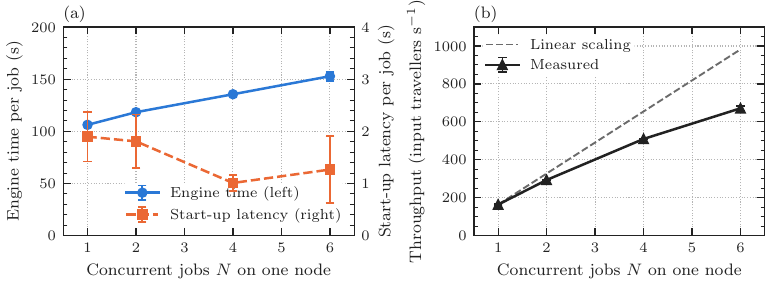}
    \caption{Concurrent execution of $N$ bus route~22 jobs on one Kubernetes node: (a)~engine time (left axis) and start-up latency (right axis) per job; (b)~aggregate throughput in input travellers per second of makespan. Error bars are sample standard deviations over all jobs of a level for engine time and start-up latency and over the five batches for throughput; the dashed line marks linear scaling of the single-job throughput.}
    \label{fig:cluster-concurrency}
\end{figure}

Table~\ref{tab:cluster} and Figure~\ref{fig:cluster-concurrency} report the results. Start-up latency ranged from 0.7 to 2.3~s over the 65 jobs and varies between batches of the same level, with level means of 1.0 to 1.9~s that do not grow with concurrency, small relative to the 106.2~s engine time of a job. Engine time per job grows from 106.2~s alone to 152.8~s with six co-located jobs (44\%); contention among the JVMs for cores and memory bandwidth is the likely cause, unconfirmed because resource utilisation was not recorded. Aggregate throughput rose 4.1-fold with six jobs, to 671 input travellers per second; all 65 jobs completed, and every invariant report passes all twelve checks. These measurements show concurrent execution and its contention cost on this node without locating its saturation point.

\section{Discussion}\label{sec:discussion}

The first research question, declarative specification with well-formedness checking before execution, is supported for the implemented SMDL subset by the tuple-level mapping and the loader's rejection of the four injected faults, one in an SMDL document and three in the transport manifest. The checks establish selected well-formedness conditions; they do not establish complete formalism conformance or behavioural conformance to DEVS execution semantics, and no mechanised proof is claimed.

The second research question names three properties, for which the evidence differs. Ephemeral execution rests on the four engine adaptations, under which the engine runs as a Kubernetes job that fetches its inputs, posts its results, and keeps no execution state beyond the run; this path was exercised functionally on the project's deployment. Scenario-level concurrency was measured with up to six jobs on one node; scale-out across nodes is untested, and no matched non-Kubernetes baseline exists for an isolation overhead. Reproducibility has three parts: reconstruction of the reported experiments is supported by the artefact bundle (Section~\ref{subsec:validation}), automatic provenance in the platform is incomplete because the image digest is not retained, and numerical repeatability was shown as identical arrival counts across repetitions on one image and node, untested across environments.

Variation of a realistic model through configuration alone at a throughput sufficient for iterative use, the third research question, is supported by the bus route~22 case study, which, within the caveats below, varied a multi-modal scenario through manifest and data changes on one model version at no less than $450\times$ real time. A structural revision through validation, publishing, execution, and reversion was not exercised, so the evidence for model evolution is the lifecycle design and the immutable registry history.

Among the systems compared in Table~\ref{tab:related_works_comparison}, the closest neighbour is DEVSML~3.0~\cite{Mittal2017}, which shares cloud-native DEVS deployment, declarative model specification, and formal DEVS semantics; the cited work reports no CI/CD lifecycle, immutable versioned artefacts, or loader-based static semantic validation before every run, which this paper adds. xDEVS~\cite{RiscoMartin2023,RiscoMartin2022} offers formal DEVS semantics and a cloud-enabled distributed architecture without a declarative model lifecycle, and the Modelica-based platform of Kumar et al.~\cite{Kumar2022} and KTWIN~\cite{Wermann2025} provide declarative twin definitions on cloud-native platforms without model CI/CD or versioning. The equilibration loop is close in spirit to MATSim~\cite{Horni2016}; it runs inside the DEVS engine, so simulation, scoring, and replanning share one event semantics and one versioned artefact.

\emph{Engine portability.} The registry, publishing worker, version pinning, SSDL scenario assembly, job dispatch, and the result-callback and lifecycle-event contracts are engine-agnostic: they treat the model as an opaque versioned artefact plus a manifest. The SMDL-to-multiPDEVS mapping and the adaptations that externalise inputs and results are engine-specific. A different DEVS engine, such as xDEVS~\cite{RiscoMartin2023} or PyPDEVS~\cite{VanTendeloo2015}, would have to implement the same four adaptations, and DEVS/DESS would need additional abstraction in the SMDL mapping; none has been hosted yet.

\emph{Deployment portability.} The Kubernetes dependency is confined to the Simulation Orchestrator, whose job renderer, service addressing, and reconciliation worker use the Kubernetes API; the Simulation API, registry, and publishing worker do not. A different container runtime therefore replaces one service; this boundary follows from the architecture, and no second runtime has been exercised.

\emph{Evaluation limits.} The evaluation covers transport only; a manufacturing or energy model would exercise continuous dynamics and the co-simulation paths (FMI and HELICS adapters of Section~\ref{subsec:mapping}) that the bus route~22 study does not. Engine performance was measured with a fixed 4~GB JVM heap; just-in-time (JIT) compiler warm-up and heap-size sensitivity were not studied. The transit demand is derived from the e-ticket validation counts and times of a single weekday, with line directions and stops assigned at random. The car demand is synthetic and inherits the temporal profile of the transit demand. Travel-time and behavioural parameters were not calibrated against observations. The point-queue link model does not capture signalised intersection dynamics, so the sub-capacity baseline does not demonstrate that the engine handles saturated traffic correctly.

\section{Conclusion and Future Work}\label{sec:conclusion}

The architecture presented in this paper makes DEVS simulation models first-class DevOps artefacts: SMDL gives them a declarative, versioned form with a tuple-level mapping to multiPDEVS. Repository CI validates their publishing manifests and publishes them as immutable versioned artefacts, the engine loader applies the static validation layers before every run, and four engine adaptations let a classical multiPDEVS engine run as a Kubernetes job that keeps no execution state beyond the run. The offline bus route~22 case study exercised this lifecycle on a model with roughly 115{,}000 atomic components; scenarios were configured through the transport manifest without changes to engine code, and each configuration passed the static validation layers before execution.

The next steps concern the pipeline itself: the loader as a registration-blocking stage of the repository pipeline, fault-detection rates over a curated corpus of model revisions, the version-evolution and regression stages of the validation strategy, a stage that gates jobs on the engine's invariant report, and measurements of the convergence of the equilibration loop and of orchestrated multi-scenario execution across several nodes. The principal application direction is to scale the case study into a public transport digital twin for Riga, composed corridor by corridor from independently versioned SMDL models, with live bus-location feeds from the operator closing the loop between the physical system and the model. Longer-term directions include behavioural validation against declared invariants and temporal properties over lifecycle events and simulation-event traces, hybrid DEVS\,/\,DESS models, automated parameter calibration in the pipeline, application studies in manufacturing and energy systems, and a SysML~v2 front end exporting to SMDL through the standard API.

\begin{backmatter}

\section*{Statements and Declarations}

\bmhead{Funding}
This research was funded by the project ``Development of the DigiTDevOps Digital Twin Development and Operation Platform'' under the European Union's Recovery and Resilience Mechanism Plan. It falls within Reform and Investment Direction 5.1: ``Increasing Productivity Through Investment in R\&D,'' specifically under Sub-action 5.1.1.r (Reform): ``Innovation Management and Motivation for Private R\&D Investment'' and Sub-action 5.1.1.2.i (Investment): ``Support Instrument for Research and Internationalisation'' (4th round). The project number is 5.1.1.2.i.0/4/24/A/CFLA/001.

\bmhead{Competing interests}
The authors have no competing interests to declare that are relevant to the content of this article.

\bmhead{Data availability}
The model and scenario files, input datasets, job and container configurations, and raw timing data are available from the corresponding author on reasonable request.

\bmhead{Code availability}
The source code of the simulation engine is not publicly available yet; the authors plan to release it as open source.


\bibliography{references}


\begin{thebibliography}{48}
\ifx \bisbn   \undefined \def \bisbn  #1{ISBN #1}\fi
\ifx \binits  \undefined \def \binits#1{#1}\fi
\ifx \bauthor  \undefined \def \bauthor#1{#1}\fi
\ifx \batitle  \undefined \def \batitle#1{#1}\fi
\ifx \bjtitle  \undefined \def \bjtitle#1{#1}\fi
\ifx \bvolume  \undefined \def \bvolume#1{\textbf{#1}}\fi
\ifx \byear  \undefined \def \byear#1{#1}\fi
\ifx \bissue  \undefined \def \bissue#1{#1}\fi
\ifx \bfpage  \undefined \def \bfpage#1{#1}\fi
\ifx \blpage  \undefined \def \blpage #1{#1}\fi
\ifx \burl  \undefined \def \burl#1{\textsf{#1}}\fi
\ifx \doiurl  \undefined \def \doiurl#1{\url{https://doi.org/#1}}\fi
\ifx \betal  \undefined \def \betal{\textit{et al.}}\fi
\ifx \binstitute  \undefined \def \binstitute#1{#1}\fi
\ifx \binstitutionaled  \undefined \def \binstitutionaled#1{#1}\fi
\ifx \bctitle  \undefined \def \bctitle#1{#1}\fi
\ifx \beditor  \undefined \def \beditor#1{#1}\fi
\ifx \bpublisher  \undefined \def \bpublisher#1{#1}\fi
\ifx \bbtitle  \undefined \def \bbtitle#1{#1}\fi
\ifx \bedition  \undefined \def \bedition#1{#1}\fi
\ifx \bseriesno  \undefined \def \bseriesno#1{#1}\fi
\ifx \blocation  \undefined \def \blocation#1{#1}\fi
\ifx \bsertitle  \undefined \def \bsertitle#1{#1}\fi
\ifx \bsnm \undefined \def \bsnm#1{#1}\fi
\ifx \bsuffix \undefined \def \bsuffix#1{#1}\fi
\ifx \bparticle \undefined \def \bparticle#1{#1}\fi
\ifx \barticle \undefined \def \barticle#1{#1}\fi
\bibcommenthead
\ifx \bconfdate \undefined \def \bconfdate #1{#1}\fi
\ifx \botherref \undefined \def \botherref #1{#1}\fi
\ifx \url \undefined \def \url#1{\textsf{#1}}\fi
\ifx \bchapter \undefined \def \bchapter#1{#1}\fi
\ifx \bbook \undefined \def \bbook#1{#1}\fi
\ifx \bcomment \undefined \def \bcomment#1{#1}\fi
\ifx \oauthor \undefined \def \oauthor#1{#1}\fi
\ifx \citeauthoryear \undefined \def \citeauthoryear#1{#1}\fi
\ifx \endbibitem  \undefined \def \endbibitem {}\fi
\ifx \bconflocation  \undefined \def \bconflocation#1{#1}\fi
\ifx \arxivurl  \undefined \def \arxivurl#1{\textsf{#1}}\fi
\csname PreBibitemsHook\endcsname

\bibitem[\protect\citeauthoryear{Grieves and Vickers}{2017}]{Grieves2017}
\begin{bchapter}
\bauthor{\bsnm{Grieves}, \binits{M.}},
\bauthor{\bsnm{Vickers}, \binits{J.}}:
\bctitle{{Digital Twin: Mitigating Unpredictable, Undesirable Emergent Behavior
  in Complex Systems}}.
In: \bbtitle{Transdisciplinary Perspectives on Complex Systems},
pp. \bfpage{85}--\blpage{113}.
\bpublisher{Springer},
\blocation{Cham}
(\byear{2017}).
\doiurl{10.1007/978-3-319-38756-7_4}
\end{bchapter}
\endbibitem

\bibitem[\protect\citeauthoryear{Tao et~al.}{2019}]{Tao2019}
\begin{barticle}
\bauthor{\bsnm{Tao}, \binits{F.}},
\bauthor{\bsnm{Zhang}, \binits{H.}},
\bauthor{\bsnm{Liu}, \binits{A.}},
\bauthor{\bsnm{Nee}, \binits{A.Y.C.}}:
\batitle{Digital twin in industry: State-of-the-art}.
\bjtitle{IEEE Transactions on Industrial Informatics}
\bvolume{15}(\bissue{4}),
\bfpage{2405}--\blpage{2415}
(\byear{2019})
\doiurl{10.1109/TII.2018.2873186}
\end{barticle}
\endbibitem

\bibitem[\protect\citeauthoryear{Zeigler et~al.}{2018}]{Zeigler2018}
\begin{bbook}
\bauthor{\bsnm{Zeigler}, \binits{B.P.}},
\bauthor{\bsnm{Muzy}, \binits{A.}},
\bauthor{\bsnm{Kofman}, \binits{E.}}:
\bbtitle{Theory of Modeling and Simulation: Discrete Event and Iterative System
  Computational Foundations},
\bedition{3rd} edn.
\bpublisher{Academic Press},
\blocation{New York}
(\byear{2018}).
\doiurl{10.1016/C2016-0-03987-6}
\end{bbook}
\endbibitem

\bibitem[\protect\citeauthoryear{Bass et~al.}{2015}]{Bass2015}
\begin{bbook}
\bauthor{\bsnm{Bass}, \binits{L.}},
\bauthor{\bsnm{Weber}, \binits{I.}},
\bauthor{\bsnm{Zhu}, \binits{L.}}:
\bbtitle{DevOps: A Software Architect's Perspective}.
\bpublisher{Addison-Wesley Professional},
\blocation{Boston, MA}
(\byear{2015})
\end{bbook}
\endbibitem

\bibitem[\protect\citeauthoryear{Niyonkuru and Wainer}{2021}]{Niyonkuru2021}
\begin{barticle}
\bauthor{\bsnm{Niyonkuru}, \binits{D.}},
\bauthor{\bsnm{Wainer}, \binits{G.}}:
\batitle{A {DEVS}-based engine for building digital quadruplets}.
\bjtitle{Simulation}
\bvolume{97}(\bissue{7}),
\bfpage{485}--\blpage{506}
(\byear{2021})
\doiurl{10.1177/00375497211003130}
\end{barticle}
\endbibitem

\bibitem[\protect\citeauthoryear{Foures et~al.}{2018}]{Foures2018}
\begin{barticle}
\bauthor{\bsnm{Foures}, \binits{D.}},
\bauthor{\bsnm{Franceschini}, \binits{R.}},
\bauthor{\bsnm{Bisgambiglia}, \binits{P.-A.}},
\bauthor{\bsnm{Zeigler}, \binits{B.P.}}:
\batitle{{multiPDEVS}: A parallel multicomponent system specification
  formalism}.
\bjtitle{Complexity}
\bvolume{2018}(\bissue{1}),
\bfpage{3751917}
(\byear{2018})
\doiurl{10.1155/2018/3751917}
\end{barticle}
\endbibitem

\bibitem[\protect\citeauthoryear{Mittal and
  Risco-Mart{\'\i}n}{2017}]{Mittal2017}
\begin{bchapter}
\bauthor{\bsnm{Mittal}, \binits{S.}},
\bauthor{\bsnm{Risco-Mart{\'\i}n}, \binits{J.L.}}:
\bctitle{{DEVSML} 3.0 stack: rapid deployment of {DEVS} farm in distributed
  cloud environment using microservices and containers}.
In: \bbtitle{Proceedings of the Symposium on Theory of Modeling \& Simulation}.
\bpublisher{Society for Modeling and Simulation International (SCS)},
\blocation{San Diego, CA, USA}
(\byear{2017}).
\doiurl{10.22360/springsim.2017.tmsdevs.043}
\end{bchapter}
\endbibitem

\bibitem[\protect\citeauthoryear{Chand et~al.}{2025}]{Chand2025}
\begin{botherref}
\oauthor{\bsnm{Chand}, \binits{S.}},
\oauthor{\bsnm{Pfannenstiel}, \binits{M.}},
\oauthor{\bsnm{Bremer}, \binits{S.}}:
Implementation of {DevOps} with {GitLab CI/CD} for the management of a large
  satellite simulation software.
Electronic Communications of the EASST
\textbf{85}
(2025)
\doiurl{10.14279/eceasst.v85.2713}
\end{botherref}
\endbibitem

\bibitem[\protect\citeauthoryear{Reiterer et~al.}{2023}]{Reiterer2023}
\begin{bchapter}
\bauthor{\bsnm{Reiterer}, \binits{S.H.}},
\bauthor{\bsnm{Schiffer}, \binits{C.}},
\bauthor{\bsnm{Schwaiger}, \binits{M.}}:
\bctitle{A graph-based meta-data model for {DevOps}: Extensions to {SSP} and
  {SysML2} and a review on the {DCP} standard}.
In: \bbtitle{Proceedings of the 15th International Modelica Conference}.
\bsertitle{Link{\"o}ping Electronic Conference Proceedings},
vol. \bseriesno{204},
pp. \bfpage{159}--\blpage{166}
(\byear{2023}).
\doiurl{10.3384/ecp204159}
\end{bchapter}
\endbibitem

\bibitem[\protect\citeauthoryear{Kreuzberger et~al.}{2023}]{Kreuzberger2023}
\begin{barticle}
\bauthor{\bsnm{Kreuzberger}, \binits{D.}},
\bauthor{\bsnm{K{\"u}hl}, \binits{N.}},
\bauthor{\bsnm{Hirschl}, \binits{S.}}:
\batitle{Machine learning operations ({MLOps}): Overview, definition, and
  architecture}.
\bjtitle{IEEE Access}
\bvolume{11},
\bfpage{31866}--\blpage{31879}
(\byear{2023})
\doiurl{10.1109/ACCESS.2023.3262138}
\end{barticle}
\endbibitem

\bibitem[\protect\citeauthoryear{Karamitsos et~al.}{2020}]{Karamitsos2020}
\begin{barticle}
\bauthor{\bsnm{Karamitsos}, \binits{I.}},
\bauthor{\bsnm{Thabit}, \binits{S.}},
\bauthor{\bsnm{Apostolopoulos}, \binits{C.}}:
\batitle{Applying {DevOps} practices of continuous automation for machine
  learning}.
\bjtitle{Information}
\bvolume{11}(\bissue{7}),
\bfpage{363}
(\byear{2020})
\doiurl{10.3390/info11070363}
\end{barticle}
\endbibitem

\bibitem[\protect\citeauthoryear{Lektauers}{2025}]{Lektauers2025}
\begin{bchapter}
\bauthor{\bsnm{Lektauers}, \binits{A.}}:
\bctitle{Towards a {DEVS}-based simulation engine for digital twin
  applications}.
In: \bbtitle{2025 Winter Simulation Conference (WSC)},
pp. \bfpage{2884}--\blpage{2895}.
\bpublisher{IEEE},
\blocation{Seattle, WA, USA}
(\byear{2025}).
\doiurl{10.1109/WSC68292.2025.11338915}
\end{bchapter}
\endbibitem

\bibitem[\protect\citeauthoryear{Hewage et~al.}{2024}]{Hewage2024}
\begin{barticle}
\bauthor{\bsnm{Hewage}, \binits{T.B.}},
\bauthor{\bsnm{Ilager}, \binits{S.}},
\bauthor{\bsnm{Rodriguez}, \binits{M.A.}},
\bauthor{\bsnm{Buyya}, \binits{R.}}:
\batitle{{CloudSim Express}: A novel framework for rapid low code simulation of
  {Cloud Computing} environments}.
\bjtitle{Software: Practice and Experience}
\bvolume{54}(\bissue{3}),
\bfpage{483}--\blpage{500}
(\byear{2024})
\doiurl{10.1002/spe.3290}
\end{barticle}
\endbibitem

\bibitem[\protect\citeauthoryear{Andreoli et~al.}{2025}]{Andreoli2025}
\begin{barticle}
\bauthor{\bsnm{Andreoli}, \binits{R.}},
\bauthor{\bsnm{Zhao}, \binits{J.}},
\bauthor{\bsnm{Cucinotta}, \binits{T.}},
\bauthor{\bsnm{Buyya}, \binits{R.}}:
\batitle{{CloudSim 7G}: An integrated toolkit for modeling and simulation of
  future generation {Cloud Computing} environments}.
\bjtitle{Software: Practice and Experience}
\bvolume{55}(\bissue{6}),
\bfpage{1041}--\blpage{1058}
(\byear{2025})
\doiurl{10.1002/spe.3413}
\end{barticle}
\endbibitem

\bibitem[\protect\citeauthoryear{Khan et~al.}{2023}]{Khan2023}
\begin{barticle}
\bauthor{\bsnm{Khan}, \binits{M.G.}},
\bauthor{\bsnm{Taheri}, \binits{J.}},
\bauthor{\bsnm{Al-Dulaimy}, \binits{A.}},
\bauthor{\bsnm{Kassler}, \binits{A.}}:
\batitle{{PerfSim}: {A Performance Simulator for Cloud Native Microservice
  Chains}}.
\bjtitle{IEEE Transactions on Cloud Computing}
\bvolume{11}(\bissue{2}),
\bfpage{1395}--\blpage{1413}
(\byear{2023})
\doiurl{10.1109/TCC.2021.3135757}
\end{barticle}
\endbibitem

\bibitem[\protect\citeauthoryear{Stoja et~al.}{2026}]{Stoja2026}
\begin{barticle}
\bauthor{\bsnm{Stoja}, \binits{S.}},
\bauthor{\bsnm{Capko}, \binits{D.}},
\bauthor{\bsnm{Jelacic}, \binits{B.}},
\bauthor{\bsnm{Vukmirovic}, \binits{S.}},
\bauthor{\bsnm{Nedic}, \binits{N.}}:
\batitle{Simulation of cloud-native microservices-based architecture for power
  applications}.
\bjtitle{Advances in Electrical and Computer Engineering}
\bvolume{26}(\bissue{1}),
\bfpage{23}--\blpage{30}
(\byear{2026})
\doiurl{10.4316/aece.2026.01003}
\end{barticle}
\endbibitem

\bibitem[\protect\citeauthoryear{Kumar et~al.}{2022}]{Kumar2022}
\begin{bchapter}
\bauthor{\bsnm{Kumar}, \binits{A.}},
\bauthor{\bsnm{Narasimhan}, \binits{A.}},
\bauthor{\bsnm{Rajendran}, \binits{T.}},
\bauthor{\bsnm{Velut}, \binits{S.}}:
\bctitle{Digital twin applications using a cloud native {Modelica} platform}.
In: \bbtitle{Proceedings of the Asian Modelica Conference 2022}.
\bsertitle{Link{\"o}ping Electronic Conference Proceedings},
vol. \bseriesno{193}.
\bpublisher{Link{\"o}ping University Electronic Press},
\blocation{Tokyo, Japan}
(\byear{2022}).
\doiurl{10.3384/ecp19375}
\end{bchapter}
\endbibitem

\bibitem[\protect\citeauthoryear{Risco-Mart{\'\i}n
  et~al.}{2023}]{RiscoMartin2023}
\begin{barticle}
\bauthor{\bsnm{Risco-Mart{\'\i}n}, \binits{J.L.}},
\bauthor{\bsnm{Mittal}, \binits{S.}},
\bauthor{\bsnm{Henares}, \binits{K.}},
\bauthor{\bsnm{Cardenas}, \binits{R.}},
\bauthor{\bsnm{Arroba}, \binits{P.}}:
\batitle{{xDEVS}: A toolkit for interoperable modeling and simulation of formal
  discrete event systems}.
\bjtitle{Software: Practice and Experience}
\bvolume{53}(\bissue{3}),
\bfpage{748}--\blpage{789}
(\byear{2023})
\doiurl{10.1002/spe.3168}
\end{barticle}
\endbibitem

\bibitem[\protect\citeauthoryear{Risco-Mart{\'\i}n
  et~al.}{2022}]{RiscoMartin2022}
\begin{barticle}
\bauthor{\bsnm{Risco-Mart{\'\i}n}, \binits{J.L.}},
\bauthor{\bsnm{Henares}, \binits{K.}},
\bauthor{\bsnm{Mittal}, \binits{S.}},
\bauthor{\bsnm{Almendras}, \binits{L.F.}},
\bauthor{\bsnm{Olcoz}, \binits{K.}}:
\batitle{A unified cloud-enabled discrete event parallel and distributed
  simulation architecture}.
\bjtitle{Simulation Modelling Practice and Theory}
\bvolume{118},
\bfpage{102539}
(\byear{2022})
\doiurl{10.1016/j.simpat.2022.102539}
\end{barticle}
\endbibitem

\bibitem[\protect\citeauthoryear{Wermann and Wickboldt}{2025}]{Wermann2025}
\begin{barticle}
\bauthor{\bsnm{Wermann}, \binits{A.G.}},
\bauthor{\bsnm{Wickboldt}, \binits{J.A.}}:
\batitle{{KTWIN}: A serverless {Kubernetes}-based {Digital Twin} platform}.
\bjtitle{Computer Networks}
\bvolume{259},
\bfpage{111095}
(\byear{2025})
\doiurl{10.1016/j.comnet.2025.111095}
\end{barticle}
\endbibitem

\bibitem[\protect\citeauthoryear{Fuller et~al.}{2020}]{Fuller2020}
\begin{barticle}
\bauthor{\bsnm{Fuller}, \binits{A.}},
\bauthor{\bsnm{Fan}, \binits{Z.}},
\bauthor{\bsnm{Day}, \binits{C.}},
\bauthor{\bsnm{Barlow}, \binits{C.}}:
\batitle{{Digital Twin: Enabling Technologies, Challenges and Open Research}}.
\bjtitle{IEEE Access}
\bvolume{8},
\bfpage{108952}--\blpage{108971}
(\byear{2020})
\doiurl{10.1109/access.2020.2998358}
\end{barticle}
\endbibitem

\bibitem[\protect\citeauthoryear{Wainer and Govind}{2024}]{Wainer2024}
\begin{barticle}
\bauthor{\bsnm{Wainer}, \binits{G.}},
\bauthor{\bsnm{Govind}, \binits{S.}}:
\batitle{100 volumes of {SIMULATION} -- 20 years of {DEVS} research}.
\bjtitle{Simulation}
\bvolume{100}(\bissue{12}),
\bfpage{1297}--\blpage{1318}
(\byear{2024})
\doiurl{10.1177/00375497241291871}
\end{barticle}
\endbibitem

\bibitem[\protect\citeauthoryear{Nutaro}{2011}]{Nutaro2011}
\begin{bbook}
\bauthor{\bsnm{Nutaro}, \binits{J.J.}}:
\bbtitle{Building Software for Simulation: Theory and Algorithms, with
  Applications in C++}.
\bpublisher{Wiley Online Library},
\blocation{Hoboken, NJ, USA}
(\byear{2011}).
\doiurl{10.1002/9780470877999}
\end{bbook}
\endbibitem

\bibitem[\protect\citeauthoryear{Wainer}{2009}]{Wainer2009}
\begin{bbook}
\bauthor{\bsnm{Wainer}, \binits{G.A.}}:
\bbtitle{Discrete-Event Modeling and Simulation: A Practitioner's Approach}.
\bpublisher{CRC Press},
\blocation{Boca Raton, FL}
(\byear{2009}).
\doiurl{10.1201/9781420053371}
\end{bbook}
\endbibitem

\bibitem[\protect\citeauthoryear{Quesnel et~al.}{2009}]{Quesnel2009}
\begin{barticle}
\bauthor{\bsnm{Quesnel}, \binits{G.}},
\bauthor{\bsnm{Duboz}, \binits{R.}},
\bauthor{\bsnm{Ramat}, \binits{{\'E}.}}:
\batitle{The {Virtual Laboratory Environment} -- {An} operational framework for
  multi-modelling, simulation and analysis of complex dynamical systems}.
\bjtitle{Simulation Modelling Practice and Theory}
\bvolume{17}(\bissue{4}),
\bfpage{641}--\blpage{653}
(\byear{2009})
\doiurl{10.1016/j.simpat.2008.11.003}
\end{barticle}
\endbibitem

\bibitem[\protect\citeauthoryear{Van~Tendeloo and
  Vangheluwe}{2015}]{VanTendeloo2015}
\begin{bchapter}
\bauthor{\bsnm{Van~Tendeloo}, \binits{Y.}},
\bauthor{\bsnm{Vangheluwe}, \binits{H.}}:
\bctitle{{PythonPDEVS}: a distributed {Parallel DEVS} simulator}.
In: \bbtitle{Proceedings of the Symposium on Theory of Modeling \& Simulation:
  {DEVS} Integrative {M\&S} Symposium}.
\bsertitle{Simulation Series},
vol. \bseriesno{47},
pp. \bfpage{91}--\blpage{98}.
\bpublisher{The Society for Modeling and Simulation International ({SCS})},
\blocation{Alexandria, VA, USA}
(\byear{2015})
\end{bchapter}
\endbibitem

\bibitem[\protect\citeauthoryear{Kim et~al.}{2009}]{Kim2009}
\begin{bchapter}
\bauthor{\bsnm{Kim}, \binits{S.}},
\bauthor{\bsnm{Sarjoughian}, \binits{H.S.}},
\bauthor{\bsnm{Elamvazhuthi}, \binits{V.}}:
\bctitle{{DEVS-Suite}: A simulator supporting visual experimentation design and
  behavior monitoring}.
In: \bbtitle{Proceedings of the 2009 Spring Simulation Multiconference}.
\bsertitle{SpringSim '09}.
\bpublisher{Society for Computer Simulation International},
\blocation{San Diego, CA, USA}
(\byear{2009}).
\bcomment{Article 161, 7 pages}
\end{bchapter}
\endbibitem

\bibitem[\protect\citeauthoryear{Camus et~al.}{2018}]{Camus2018}
\begin{barticle}
\bauthor{\bsnm{Camus}, \binits{B.}},
\bauthor{\bsnm{Paris}, \binits{T.}},
\bauthor{\bsnm{Vaubourg}, \binits{J.}},
\bauthor{\bsnm{Presse}, \binits{Y.}},
\bauthor{\bsnm{Bourjot}, \binits{C.}},
\bauthor{\bsnm{Ciarletta}, \binits{L.}},
\bauthor{\bsnm{Chevrier}, \binits{V.}}:
\batitle{Co-simulation of cyber-physical systems using a {DEVS} wrapping
  strategy in the {MECSYCO} middleware}.
\bjtitle{Simulation}
\bvolume{94}(\bissue{12}),
\bfpage{1099}--\blpage{1127}
(\byear{2018})
\doiurl{10.1177/0037549717749014}
\end{barticle}
\endbibitem

\bibitem[\protect\citeauthoryear{Vanommeslaeghe
  et~al.}{2024}]{Vanommeslaeghe2024}
\begin{bchapter}
\bauthor{\bsnm{Vanommeslaeghe}, \binits{Y.}},
\bauthor{\bsnm{Van~Acker}, \binits{B.}},
\bauthor{\bsnm{Denil}, \binits{J.}},
\bauthor{\bsnm{De~Meulenaere}, \binits{P.}}:
\bctitle{Integrating {DEVS} and {FMI} 3.0 for the simulated deployment of
  embedded applications}.
In: \bbtitle{2024 Annual Modeling and Simulation Conference ({ANNSIM})},
pp. \bfpage{1}--\blpage{13}.
\bpublisher{IEEE},
\blocation{Washington, D.C., DC, USA}
(\byear{2024}).
\doiurl{10.23919/ANNSIM61499.2024.10732719}
\end{bchapter}
\endbibitem

\bibitem[\protect\citeauthoryear{Calheiros et~al.}{2011}]{Calheiros2011}
\begin{barticle}
\bauthor{\bsnm{Calheiros}, \binits{R.N.}},
\bauthor{\bsnm{Ranjan}, \binits{R.}},
\bauthor{\bsnm{Beloglazov}, \binits{A.}},
\bauthor{\bsnm{De~Rose}, \binits{C.A.F.}},
\bauthor{\bsnm{Buyya}, \binits{R.}}:
\batitle{{CloudSim}: a toolkit for modeling and simulation of cloud computing
  environments and evaluation of resource provisioning algorithms}.
\bjtitle{Software: Practice and Experience}
\bvolume{41}(\bissue{1}),
\bfpage{23}--\blpage{50}
(\byear{2011})
\doiurl{10.1002/spe.995}
\end{barticle}
\endbibitem

\bibitem[\protect\citeauthoryear{{da Gi{\~a}o} et~al.}{2024}]{daGiao2024}
\begin{botherref}
\oauthor{\bsnm{{da Gi{\~a}o}}, \binits{H.}},
\oauthor{\bsnm{Flores}, \binits{A.}},
\oauthor{\bsnm{Pereira}, \binits{R.}},
\oauthor{\bsnm{Cunha}, \binits{J.}}:
Chronicles of {CI/CD}: A Deep Dive into its Usage Over Time.
\url{https://arxiv.org/abs/2402.17588}.
arXiv:2402.17588
(2024)
\end{botherref}
\endbibitem

\bibitem[\protect\citeauthoryear{Capizzi et~al.}{2020}]{Capizzi2020}
\begin{bchapter}
\bauthor{\bsnm{Capizzi}, \binits{A.}},
\bauthor{\bsnm{Distefano}, \binits{S.}},
\bauthor{\bsnm{Mazzara}, \binits{M.}}:
\bctitle{From {DevOps} to {DevDataOps}: Data management in {DevOps} processes}.
In: \bbtitle{Software Engineering Aspects of Continuous Development and New
  Paradigms of Software Production and Deployment},
pp. \bfpage{52}--\blpage{62}.
\bpublisher{Springer},
\blocation{Cham}
(\byear{2020}).
\doiurl{10.1007/978-3-030-39306-9_4}
\end{bchapter}
\endbibitem

\bibitem[\protect\citeauthoryear{Colantoni et~al.}{2020}]{Colantoni2020}
\begin{bchapter}
\bauthor{\bsnm{Colantoni}, \binits{A.}},
\bauthor{\bsnm{Berardinelli}, \binits{L.}},
\bauthor{\bsnm{Wimmer}, \binits{M.}}:
\bctitle{{DevOpsML}: towards modeling {DevOps} processes and platforms}.
In: \bbtitle{Proceedings of the 23rd ACM/IEEE International Conference on Model
  Driven Engineering Languages and Systems: Companion Proceedings},
pp. \bfpage{1}--\blpage{10}.
\bpublisher{ACM},
\blocation{New York, NY, USA}
(\byear{2020}).
\doiurl{10.1145/3417990.3420203}
\end{bchapter}
\endbibitem

\bibitem[\protect\citeauthoryear{Subramanya et~al.}{2022}]{Subramanya2022}
\begin{barticle}
\bauthor{\bsnm{Subramanya}, \binits{R.}},
\bauthor{\bsnm{Sierla}, \binits{S.}},
\bauthor{\bsnm{Vyatkin}, \binits{V.}}:
\batitle{From {DevOps} to {MLOps}: Overview and application to electricity
  market forecasting}.
\bjtitle{Applied Sciences}
\bvolume{12}(\bissue{19}),
\bfpage{9851}
(\byear{2022})
\doiurl{10.3390/app12199851}
\end{barticle}
\endbibitem

\bibitem[\protect\citeauthoryear{Kamburjan et~al.}{2021}]{Kamburjan2021}
\begin{bchapter}
\bauthor{\bsnm{Kamburjan}, \binits{E.}},
\bauthor{\bsnm{Klungre}, \binits{V.N.}},
\bauthor{\bsnm{Schlatte}, \binits{R.}},
\bauthor{\bsnm{Johnsen}, \binits{E.B.}},
\bauthor{\bsnm{Giese}, \binits{M.}}:
\bctitle{Programming and debugging with semantically lifted states}.
In: \bbtitle{18th European Semantic Web Conference (ESWC 2021)}.
\bsertitle{LNCS},
vol. \bseriesno{12731},
pp. \bfpage{126}--\blpage{142}.
\bpublisher{Springer},
\blocation{Cham}
(\byear{2021}).
\doiurl{10.1007/978-3-030-77385-4_8}
\end{bchapter}
\endbibitem

\bibitem[\protect\citeauthoryear{{Object Management
  Group}}{2025}]{OMG2025sysml}
\begin{botherref}
\oauthor{\bsnm{{Object Management Group}}}:
{OMG Systems Modeling Language (SysML)}, Version 2.0.
\url{https://www.omg.org/spec/SysML/2.0}.
Formal specification; published together with {Kernel Modeling Language
  (KerML)} Version 1.0 and {Systems Modeling API and Services} Version 1.0
(2025)
\end{botherref}
\endbibitem

\bibitem[\protect\citeauthoryear{{Microsoft Azure}}{2026}]{Microsoft2026}
\begin{botherref}
\oauthor{\bsnm{{Microsoft Azure}}}:
{Digital Twins Definition Language} ({DTDL}).
\url{https://github.com/Azure/opendigitaltwins-dtdl}.
v4, accessed April 2026
(2026)
\end{botherref}
\endbibitem

\bibitem[\protect\citeauthoryear{Kapos et~al.}{2014}]{Kapos2014}
\begin{barticle}
\bauthor{\bsnm{Kapos}, \binits{G.-D.}},
\bauthor{\bsnm{Dalakas}, \binits{V.}},
\bauthor{\bsnm{Nikolaidou}, \binits{M.}},
\bauthor{\bsnm{Anagnostopoulos}, \binits{D.}}:
\batitle{An integrated framework for automated simulation of {SysML} models
  using {DEVS}}.
\bjtitle{Simulation}
\bvolume{90}(\bissue{6}),
\bfpage{717}--\blpage{744}
(\byear{2014})
\doiurl{10.1177/0037549714533842}
\end{barticle}
\endbibitem

\bibitem[\protect\citeauthoryear{Blochwitz et~al.}{2012}]{Blochwitz2012}
\begin{bchapter}
\bauthor{\bsnm{Blochwitz}, \binits{T.}},
\bauthor{\bsnm{Otter}, \binits{M.}},
\bauthor{\bsnm{Akesson}, \binits{J.}},
\bauthor{\bsnm{Arnold}, \binits{M.}},
\bauthor{\bsnm{Clauss}, \binits{C.}},
\bauthor{\bsnm{Elmqvist}, \binits{H.}},
\bauthor{\bsnm{Friedrich}, \binits{M.}},
\bauthor{\bsnm{Junghanns}, \binits{A.}},
\bauthor{\bsnm{Mauss}, \binits{J.}},
\bauthor{\bsnm{Neumerkel}, \binits{D.}},
\bauthor{\bsnm{Olsson}, \binits{H.}},
\bauthor{\bsnm{Viel}, \binits{A.}}:
\bctitle{Functional mockup interface 2.0: The standard for tool independent
  exchange of simulation models}.
In: \bbtitle{Proceedings of the 9th International Modelica Conference},
vol. \bseriesno{76},
pp. \bfpage{173}--\blpage{184}.
\bpublisher{Link{\"o}ping University Electronic Press},
\blocation{Munich, Germany}
(\byear{2012}).
\doiurl{10.3384/ecp12076173}
\end{bchapter}
\endbibitem

\bibitem[\protect\citeauthoryear{Hardy et~al.}{2024}]{Hardy2024}
\begin{barticle}
\bauthor{\bsnm{Hardy}, \binits{T.D.}},
\bauthor{\bsnm{Palmintier}, \binits{B.}},
\bauthor{\bsnm{Top}, \binits{P.L.}},
\bauthor{\bsnm{Krishnamurthy}, \binits{D.}},
\bauthor{\bsnm{Fuller}, \binits{J.C.}}:
\batitle{{HELICS}: A co-simulation framework for scalable multi-domain modeling
  and analysis}.
\bjtitle{IEEE Access}
\bvolume{12},
\bfpage{24325}--\blpage{24347}
(\byear{2024})
\doiurl{10.1109/ACCESS.2024.3363615}
\end{barticle}
\endbibitem

\bibitem[\protect\citeauthoryear{Holcombe}{1988}]{Holcombe1988}
\begin{barticle}
\bauthor{\bsnm{Holcombe}, \binits{M.}}:
\batitle{{X}-machines as a basis for dynamic system specification}.
\bjtitle{Software Engineering Journal}
\bvolume{3}(\bissue{2}),
\bfpage{69}--\blpage{76}
(\byear{1988})
\doiurl{10.1049/sej.1988.0009}
\end{barticle}
\endbibitem

\bibitem[\protect\citeauthoryear{Lektauers and Merkuryev}{2014}]{Lektauers2014}
\begin{bchapter}
\bauthor{\bsnm{Lektauers}, \binits{A.}},
\bauthor{\bsnm{Merkuryev}, \binits{Y.}}:
\bctitle{{DEVS}-based interactive geosimulation framework for public transport
  analysis and planning}.
In: \bbtitle{2nd International Workshop on Simulation for Energy, Sustainable
  Development and Environment (SESDE 2014)},
pp. \bfpage{50}--\blpage{58}.
\bpublisher{{Dime University of Genoa}},
\blocation{Bordeaux, France}
(\byear{2014})
\end{bchapter}
\endbibitem

\bibitem[\protect\citeauthoryear{{OpenStreetMap
  contributors}}{2026}]{OpenStreetMap2026}
\begin{botherref}
\oauthor{\bsnm{{OpenStreetMap contributors}}}:
{OpenStreetMap}.
\url{https://www.openstreetmap.org}.
Extract for Latvia of 6 August 2026
(2026)
\end{botherref}
\endbibitem

\bibitem[\protect\citeauthoryear{{R{\={\i}}gas
  Satiksme}}{2026a}]{RigasSatiksme2026a}
\begin{botherref}
\oauthor{\bsnm{{R{\={\i}}gas Satiksme}}}:
{Public transport route schedules (GTFS feed)}.
\url{https://data.gov.lv/dati/lv/dataset/marsrutu-saraksti-rigas-satiksme-sabiedriskajam-transportam}.
Latvian Open Data Portal, July 2026 feed
(2026)
\end{botherref}
\endbibitem

\bibitem[\protect\citeauthoryear{{R{\={\i}}gas
  Satiksme}}{2026b}]{RigasSatiksme2026b}
\begin{botherref}
\oauthor{\bsnm{{R{\={\i}}gas Satiksme}}}:
{E-ticket validation data for public transport}.
\url{https://data.gov.lv/dati/lv/dataset/e-talonu-validaciju-dati-rigas-satiksme-sabiedriskajos-transportlidzeklos}.
Latvian Open Data Portal, June 2026 data
(2026)
\end{botherref}
\endbibitem

\bibitem[\protect\citeauthoryear{El-Geneidy et~al.}{2014}]{ElGeneidy2014}
\begin{barticle}
\bauthor{\bsnm{El-Geneidy}, \binits{A.}},
\bauthor{\bsnm{Grimsrud}, \binits{M.}},
\bauthor{\bsnm{Wasfi}, \binits{R.}},
\bauthor{\bsnm{T{\'e}treault}, \binits{P.}},
\bauthor{\bsnm{Surprenant-Legault}, \binits{J.}}:
\batitle{New evidence on walking distances to transit stops: Identifying
  redundancies and gaps using variable service areas}.
\bjtitle{Transportation}
\bvolume{41}(\bissue{1}),
\bfpage{193}--\blpage{210}
(\byear{2014})
\doiurl{10.1007/s11116-013-9508-z}
\end{barticle}
\endbibitem

\bibitem[\protect\citeauthoryear{Horni et~al.}{2016}]{Horni2016}
\begin{bbook}
\bauthor{\bsnm{Horni}, \binits{A.}},
\bauthor{\bsnm{Nagel}, \binits{K.}},
\bauthor{\bsnm{Axhausen}, \binits{K.W.}}:
\bbtitle{The Multi-agent Transport Simulation {MATSim}}.
\bpublisher{Ubiquity Press},
\blocation{London}
(\byear{2016}).
\doiurl{10.5334/baw}
\end{bbook}
\endbibitem

\bibitem[\protect\citeauthoryear{Charypar and Nagel}{2005}]{Charypar2005}
\begin{barticle}
\bauthor{\bsnm{Charypar}, \binits{D.}},
\bauthor{\bsnm{Nagel}, \binits{K.}}:
\batitle{Generating complete all-day activity plans with genetic algorithms}.
\bjtitle{Transportation}
\bvolume{32}(\bissue{4}),
\bfpage{369}--\blpage{397}
(\byear{2005})
\doiurl{10.1007/s11116-004-8287-y}
\end{barticle}
\endbibitem

\end{thebibliography}

\end{backmatter}

\end{document}